\documentclass[aps, prd, superscriptaddress, preprintnumbers]{revtex4}

\usepackage{graphicx}
\usepackage{amsmath}
\usepackage{amssymb}
\usepackage{color}
\usepackage{bm}
\usepackage[normalem]{ulem}

\newcommand{\be}{\begin{equation}}
\newcommand{\ee}{\end{equation}}
\newcommand{\ba}{\begin{eqnarray}}
\newcommand{\ea}{\end{eqnarray}}

\newcommand{\sign}{\,\mbox{sign}}

\definecolor{red}{rgb}{0.7,0,0}
\definecolor{green}{rgb}{0,0.5,0}

\begin{document}

\title{Chiral anomaly, dimensional reduction, and magnetoresistivity of Weyl and Dirac semimetals}
\date{February 21, 2014}

\author{E. V. Gorbar}
\affiliation{Department of Physics, Taras Shevchenko National Kiev University, Kiev, 03680, Ukraine}
\affiliation{Bogolyubov Institute for Theoretical Physics, Kiev, 03680, Ukraine}

\author{V. A. Miransky}
\affiliation{Department of Applied Mathematics, Western University, London, Ontario N6A 5B7, Canada}

\author{I. A. Shovkovy}
\affiliation{School of Letters and Sciences, Arizona State University, Mesa, Arizona 85212, USA}
    
\begin{abstract}
By making use of the Kubo formula, we calculate the conductivity of Dirac and Weyl semimetals 
in a magnetic field. We find that the longitudinal (along the direction of the magnetic field) 
magnetoresistivity is negative at sufficiently large magnetic fields for {\it both} Dirac and Weyl 
semimetals. The physical reason of this phenomenon is intimately connected with the dimensional 
spatial reduction $3 \to 1$ in the dynamics of the lowest Landau level. The off-diagonal component 
of the transverse (with respect to the direction of the magnetic field) conductivity in Weyl semimetals 
contains an anomalous contribution directly proportional to the momentum space separation between 
the Weyl nodes. This contribution comes exclusively from the lowest Landau level and, as expected, 
is independent of the temperature, chemical potential, and magnetic field. The other part of the 
off-diagonal conductivity is the same as in Dirac semimetals and is connected with a nonzero 
density of charge carriers. The signatures for experimental distinguishing Weyl semimetals 
from Dirac ones through the measurements of conductivity are discussed.
\end{abstract}

\maketitle

\section{Introduction}

The discovery of graphene \cite{graphene}, whose quasiparticle excitations are described 
by the two-dimensional massless Dirac equation, drew a lot of attention to the unique electronic 
and transport properties of materials with a relativistic-like electron spectrum. As a result, 
materials with an approximate three-dimensional (3D) Dirac electron spectrum also moved 
to the forefront of theoretical and experimental studies. Historically, bismuth is the first condensed 
matter material in which the electron states near the $L$ point in the Brillouin zone are described 
by the 3D massive Dirac equation \cite{Cohen,Wolff,Falkovsky,Edelman}. It is also known that the 
corresponding value of the Dirac mass decreases when bismuth is doped with a small amount of
antimony. Moreover, such an alloy Bi$_{1-x}$Sb$_x$ with the antimony concentration of about 
$x \approx 0.03$ becomes a semimetal with massless Dirac fermions \cite{Lenoir,Teo}.

Although other materials with 3D Dirac fermions can be obtained by fine tuning the strength of 
the spin-orbital coupling or chemical composition \cite{Murakami,Zhang,Xu,Sato,Das2013}, it is difficult to 
control such realizations. An interesting idea was expressed recently in Ref.~\cite{Young}, 
where it was shown that the formation of Dirac points can be protected by a crystal symmetry, and
metastable $\beta$-cristobalite BiO$_2$ was suggested as a specific example of a massless 
Dirac material. Later, by using first-principles calculations and effective model analysis, the authors 
of Refs.~\cite{Wang,Weng} predicted that $A_3$Bi ($A$ = Na, K, Rb) and Cd$_3$As$_2$ should
be Dirac semimetals with bulk 3D Dirac points protected by crystal symmetry. The experimental
discovery of the 3D Dirac fermions in Na$_3$Bi and Cd$_3$As$_2$ was recently reported in 
Refs.~\cite{Liu} and \cite{Neupane,Borisenko}, respectively. The Dirac nature 
of the quasiparticles was confirmed by investigating the electronic structure of these materials 
with angle-resolved photoemission spectroscopy.
 
The Dirac four-component spinor is composed of two (i.e., left-handed and right-handed) 
two-component fermions. The latter are described by the Weyl equation of the corresponding 
chirality. If the requirement of inversion or time-reversal symmetry is relaxed, the degeneracy 
of the dispersion relations of the left- and right-handed Weyl modes can be lifted, 
transforming the Dirac semimetal into a Weyl one. While pyrochlore iridates \cite{Wan},
as well as some heterostructures of topological and normal insulators \cite{Burkov1}, are 
conjectured to be Weyl semimetals (for a review, see Refs.~\cite{Hook,Turner,Vafek}), 
no material at present is experimentally proved to be a Weyl semimetal.
Since magnetic field breaks time reversal symmetry, one may engineer a Weyl semimetal from 
a Dirac one by applying the external magnetic field. One such mechanism was 
originally described in the context of high-energy physics some time ago\cite{GMS2009} 
and was applied to studies of Dirac and Weyl semimetals in Ref.~\cite{engineering}.
It is expected that the same mechanism can be realized in the newly discovered 
3D Dirac semimetals Na$_3$Bi and Cd$_3$As$_2$\cite{Liu,Neupane,Borisenko} 
(in addition to the magnetic field, a necessary condition 
for this mechanism to operate is a nonzero density of charge carriers).
 
Negative longitudinal magnetoresistivity in Weyl semimetals \cite{Nielsen,Aji,Son} 
is a consequence of the chiral anomaly \cite{anomaly} and is considered in the literature as a 
fingerprint of a Weyl semimetal phase. It is noticeable that in a magnetic field the chiral 
anomaly is generated entirely on the lowest Landau level (LLL) \cite{Ambjorn}. In particular, 
the anomaly is responsible for pumping the electrons between the nodes of opposite chirality 
at a rate proportional to the scalar product of the applied electric and magnetic fields 
$\mathbf{E}\cdot\mathbf{B}$. Recently, a negative longitudinal magnetoresistivity\cite{footnote1} 
was observed in Bi$_{1-x}$Sb$_x$ alloy with $x \approx 0.03$ in moderately strong 
magnetic fields \cite{Kim1307.6990} and was interpreted as an experimental signature of
the presence of a Weyl semimetal phase, where a single Dirac point splits into two Weyl 
nodes with opposite chiralities and the separation between the nodes in momentum space is 
proportional to the applied field. As we will show in this study, however, the observation 
of the negative longitudinal magnetoresistivity is also expected in Dirac semimetals. 
Therefore, negative magnetoresistivity alone may not be sufficient to unambiguously 
distinguish between the Dirac and Weyl semimetals. Note that a nonlocal transport 
can be another way of probing the chiral anomaly in Weyl semimetals \cite{Abanin}.

In Refs.~\cite{Nielsen,Aji,Son}, the magnetoresistivity in Weyl semimetals was 
studied by using the semiclassical Boltzmann kinetic equation. Since negative longitudinal 
magnetoresistivity is one of the key characteristics of Weyl semimetals intimately connected 
with the chiral anomaly, in this paper we derive magnetoresistivity using the microscopic 
Kubo formalism, which takes into account quantum effects. (In a special class of gapless 
Dirac semiconductors with a small carrier concentration, transverse magnetoresistivity 
was previously studied in Ref.~\cite{Abrikosov}.) We found that the negative 
longitudinal magnetoresistivity takes place not only in Weyl semimetals, but also in 
Dirac ones. 

As we argue in Sec.~\ref{sec:longitudinal}, the origin of the negative magnetoresistivity 
is intimately connected with the spatial dimensional reduction $3 \to 1$ in the low-energy 
dynamics dominated by the LLL. Such a dimensional reduction is a universal phenomenon, 
taking place in the dynamics of charged fermions in a magnetic field \cite{reduction}. The 
low-energy quasiparticles are described by the spin-polarized LLL states and effectively 
have one-dimensional dispersion relations, which depend only on the longitudinal 
momentum $k_3$ and do not contain the magnetic field at all [see Eq.~(\ref{LLLprop})]. 
The physics behind this phenomenon is the following. As is well known, the 
transverse momenta $k_1$ and $k_2$ are not good quantum numbers for quasiparticles 
in a magnetic field. In the dispersion relations, such momenta are replaced by a single 
discrete quantum number $n$, labeling the Landau levels (which have a degeneracy 
proportional to the value of the magnetic field).

The consequences of the dimensional reduction are rather dramatic in the case 
of relativistic-like massless fermions because of their chiral nature: such fermions disperse 
only one way in the longitudinal direction for each chirality \cite{Qi}. The existence of massless 
chiral fermions and their high degeneracy in the presence of a magnetic field are topologically 
protected by the index theorem \cite{AharonovCasher1979}. We find that it is this unique nature 
of the low-energy states that is responsible for the main contribution (growing linearly with 
the field) to the longitudinal conductivity in Dirac/Weyl semimetals. In fact, as we will see 
in the following, the special nature of the LLL plays a profound role also in the 
anomalous Hall contribution to the transverse conductivity.

Finally, we would like to mention that electric transport in Weyl semimetals in the absence 
of magnetic field was studied in Refs.~\cite{Hosur,Rosenstein}. The magneto-optical 
conductivity of Weyl semimetals was investigated in Ref.~\cite{Ashby}. Recent 
developments in transport phenomena in Weyl semimetals are reviewed in 
Ref.~\cite{Qi} focusing on signatures connected with the chiral anomaly. 

The paper is organized as follows. The model is described and the notations are 
introduced in Sec.~\ref{sec:Model}. The quasiparticle propagator and the spectral 
function are derived in Sec.~\ref{sec:Propagator}. In Sec.~\ref{sec:KuboFormula}, 
the general expression for the conductivity in the Kubo formalism is obtained. 
The longitudinal and transverse components of the conductivity are calculated 
in Secs.~\ref{sec:longitudinal} and \ref{sec:transverse}, respectively. The results 
are summarized and the conclusion is given in Sec.~\ref{sec:Conclusion}.
For convenience, throughout this paper, we set $\hbar=1$.

\section{Model}
\label{sec:Model}

The low-energy Hamiltonian of a Weyl semimetal in an external magnetic field is given by
\begin{equation}
H^{\rm (W)}=H^{\rm (W)}_0+H_{\rm int},
\label{Hamiltonian-model-Weyl}
\end{equation}
where $H_{\rm int}$ is the electron-electron interaction Hamiltonian (since for the 
rest of this study the explicit form of $H_{\rm int}$ is not important, we do not write 
it here) and
\begin{equation}
H^{\rm (W)}_0=\int d^3r \left[\,v_F \psi^{\dagger} (\mathbf{r})\left( 
\begin{array}{cc} \bm{\sigma}\cdot(-i\bm{\nabla}+e\mathbf{A}/c -\mathbf{b}) +b_0 & 0\\ 0 & 
-\bm{\sigma}\cdot(-i\bm{\nabla}+e\mathbf{A}/c+\mathbf{b}) - b_0\end{array} 
\right)\psi(\mathbf{r})-\mu\, \psi^{\dagger} (\mathbf{r})\psi(\mathbf{r})
\right]
\label{free-Hamiltonian}
\end{equation}
is the Hamiltonian of the free theory, which describes two Weyl nodes of opposite (as required 
by the Nielsen--Ninomiya theorem \cite{Nielsen}) chirality separated by vector $2\mathbf{b}$ in 
momentum space and by $2b_0$ in energy. In the above Hamiltonian, we used the following notation: $v_F$ is the Fermi 
velocity, $\mathbf{A}$ is the vector potential, which describes a constant magnetic field, $c$ is the speed of light, 
$\mu$ is the chemical potential, and $\bm{\sigma}=(\sigma_x,\sigma_y,\sigma_z)$ are Pauli matrices. 
In the special case when $\mathbf{b}=0$ and $b_0=0$, the Hamiltonian (\ref{free-Hamiltonian}) describes a Dirac 
semimetal. Note that here we consider the simplest example of a Weyl semimetal with a single pair 
of Weyl nodes, but the generalization to a larger number of Weyl nodes is straightforward. 

We would like to remind that while the momentum shift $2\mathbf{b}$ is odd under the 
time reversal symmetry, the energy shift $2b_0$ is odd under the inversion symmetry (parity).
The experimentally discovered 
3D semimetals, mentioned in the Introduction, are all Dirac type semimetals that preserve both time 
reversal and parity. When a Weyl semimetal is produced from a Dirac one by applying an external magnetic 
field, parity will remain preserved, unlike time reversal, which is explicitly broken by the magnetic field. 
In this work, we consider only this type of Weyl semimetals. In the most general case, 
on the other hand, the Weyl points can be shifted in energy. The effect of such a shift is not 
immediately obvious because of the anomaly-related contributions that need a very careful analysis.
Some of the subtleties (although in a slightly different context) have been discussed in Refs.~\cite{Franz}
and \cite{Basar:2013iaa}. This type of Weyl systems are beyond the scope of this paper and will be 
considered elsewhere.

In the case when a Dirac semimetal is formed in a multilayer heterostructure, composed of alternating 
layers of topological and normal insulator materials without magnetic impurities, a nonzero magnetic field 
will turn it into a Weyl semimetal via the Zeeman interaction \cite{Hook}. The corresponding induced 
separation between the Weyl nodes in momentum space is determined by $\mathbf{b}=-g\mu_B\mathbf{B}$, 
where $\mathbf{B}$ is the magnetic field and $g$ is the spin $g$-factor. Note that the typical values of 
the $g$-factor in topological insulators are rather large, $g\simeq 50$ \cite{Hook}. In bismuth, on the 
other hand, the spin interaction with the magnetic field is already accounted for by the Dirac equation
of the low-energy effective theory\cite{Wolff} (recall that the spin $g$ factor 
is $g = 2$ in the Dirac equation). The same is true also for the Bi$_{1-x}$Sb$_x$ alloy with a 
low concentration of antimony \cite{ChuKao}, as well as for the recently discovered 3D Dirac semimetals 
Na$_3$Bi and Cd$_3$As$_2$ \cite{Liu,Neupane,Borisenko}. In all of these materials, therefore, there is 
no additional Zeeman interaction that would be able to generate the separation in momentum space
between the left- and right-handed modes. However, as argued in 
Ref.~\cite{engineering}, there is a different mechanism that transforms Dirac semimetals 
of this type into Weyl semimetals. The new mechanism was originally proposed in the context 
of high-energy physics \cite{GMS2009}. It is driven by the electron-electron interaction in matter 
with a nonzero density of charge carriers and has a subtle connection with the chiral anomaly.
The dynamically induced chiral shift is directed along the magnetic field and 
its magnitude is determined by the quasiparticle charge density,  the strength of the magnetic 
field, and the strength of the interaction.

Before proceeding with the analysis, we find it very convenient to introduce 
the four-dimensional Dirac matrices in the chiral representation:
\begin{equation}
\gamma^0 = \left( \begin{array}{cc} 0 & -I\\ -I & 0 \end{array} \right),\qquad
\bm{\gamma} = \left( \begin{array}{cc} 0& \bm{\sigma} \\  - \bm{\sigma} & 0 \end{array} \right),
\label{Dirac-matrices}
\end{equation}
where $I$ is the two-dimensional unit matrix. In this notation, the free Weyl Hamiltonian 
(\ref{free-Hamiltonian}) for $b_0=0$ takes the following form:
\begin{equation}
H^{\rm (W)}_0 = \int d^3 r\, \bar{\psi} (\mathbf{r})\left[
-i v_F (\bm{\gamma}\cdot (\bm{\nabla}+ie\mathbf{A}))-(\mathbf{b}\cdot \bm{\gamma})\gamma^5-\mu\gamma^0 
\right]\psi(\mathbf{r}) ,
\label{free-Hamiltonian-Weyl-rel}
\end{equation}
where, by definition, $\bar{\psi} \equiv \psi^{\dagger}\gamma^0$ is the Dirac conjugate spinor field 
and the matrix $\gamma^5$ is
\begin{equation}
\gamma^5 \equiv i\gamma^0\gamma^1\gamma^2\gamma^3 
= \left( \begin{array}{cc} I & 0\\ 0 & -I \end{array} \right).
\end{equation}
As it is clear from the first term in the free Hamiltonian (\ref{free-Hamiltonian}), the 
eigenvalues of $\gamma^5$ correspond to the node (chirality) degrees of freedom.

\section{Propagator and spectral function}
\label{sec:Propagator}

The inverse propagator in a Weyl semimetal can be written in the following form:
\begin{equation}
iG^{-1}(u,u^\prime)= \left[(i\partial_t+\mu )\gamma^0 - v_F(\bm{\pi}\cdot\bm{\gamma}) 
+v_F (\mathbf{b} \cdot \bm{\gamma})\gamma^5\right]\delta^{4}(u- u^\prime),
\label{ginverse}
\end{equation}
where $u=(t, \mathbf{r})$ and $\bm{\pi} \equiv -i \bm{\nabla} + e\mathbf{A}/c$ is the canonical 
momentum. In order to derive the propagator $G(u,u^\prime)$ in the Landau-level representation, 
we invert $G^{-1}(u,u^\prime)$ in Eq.~(\ref{ginverse}) by using the approach described in 
Appendix~\ref{AppA} of Ref.~\cite{Gorbar:2011ya}. The result takes the following form: 
\begin{eqnarray}
G(u,u^\prime) &=& e^{i\Phi(\mathbf{r}_\perp,\mathbf{r}_\perp^\prime)}\bar{G}(u-u^\prime), \\
\bar{G}(t-t^\prime;\mathbf{r}-\mathbf{r}^\prime)&=& \int \frac{d\omega d^3\mathbf{k}}{(2\pi)^4}
e^{-i\omega (t-t^\prime)+i\mathbf{k}\cdot(\mathbf{r}-\mathbf{r}^\prime)}
\bar{G}(\omega;\mathbf{k}),
\end{eqnarray}
where $\Phi(\mathbf{r}_\perp,\mathbf{r}_\perp^\prime)=-eB(x+y^{\prime})(x-y^{\prime})/2$ 
is the Schwinger phase \cite{Schwinger} for the vector potential in the Landau gauge 
$\mathbf{A}=(0,Bx,0)$, which describes the magnetic field $\mathbf{B}$ that points in 
the $+z$ direction. The propagator is described by two separate Weyl node contributions, i.e.,
\begin{equation}
\bar{G}(\omega;\mathbf{k})= \sum_{\chi=\pm}\bar{G}^{(\chi)}(\omega;\mathbf{k}) {\cal P}_{5}^{(\chi)},
\end{equation}
where the chiral shift is assumed to be along the direction of the magnetic field, i.e.,  
$\mathbf{b}=(0,0,b)$, ${\cal P}_{5}^{(\chi)}\equiv \frac12 \left(1+ \chi\gamma^5\right)$ 
are the Weyl node (chirality) projectors, and
\begin{eqnarray}
\bar{G}^{(\chi)}(\omega;\mathbf{k}) 
&=& ie^{-k_\perp^2 l^{2}}\sum_{\lambda=\pm} \sum_{n=0}^{\infty} 
 \frac{(-1)^n}{E_n^{(\chi)}} \Bigg\{ \left[ E_n^{(\chi)} \gamma^{0}  -\lambda v_F (k_3-\chi b)\gamma^3 \right]
\left[{\cal P}_{-}L_n\left(2 k_\perp^2 l^{2}\right)
-{\cal P}_{+}L_{n-1}\left(2 k_\perp^2 l^{2}\right)\right] \nonumber\\
&& +  2\lambda v_F(\mathbf{k}_\perp\cdot\bm{\gamma}_\perp) L_{n-1}^1\left(2 k_\perp^2 l^{2}\right)
\Bigg\}\frac{1}{ \omega+\mu   - \lambda E_n^{(\chi)} }.
\label{full-propagator}
\end{eqnarray}
Here $L_{n}^\alpha(z)$ are the generalized Laguerre polynomials,
$E_n^{(\chi)}=v_F \sqrt{(k_3-\chi b)^2 + 2 n |eB|/c}$ is the energy in the $n$th Landau level, 
$\mathbf{k}_\perp=(k^1,k^2)$ is the transverse pseudomomentum,  
${\cal P}_{\pm}\equiv \frac12 \left(1\pm i s_\perp \gamma^1\gamma^2\right)$ are spin 
[or pseudospin if the Pauli matrices in the free Hamiltonian (\ref{free-Hamiltonian}) are pseudospin 
matrices] projectors, and $l=\sqrt{c/|eB|}$ is the magnetic length. By definition, 
$s_\perp=\sign (eB)$ and $L_{-1}^\alpha \equiv 0$. 

The spectral function is given by the difference of the advanced and retarded propagators, i.e.,
\begin{equation}
A(\omega;\mathbf{k}) =\frac{1}{2\pi i}\left[
 \bar{G}_{\mu=0}(\omega-i 0 ;\mathbf{k}) 
-\bar{G}_{\mu=0}(\omega+i 0 ;\mathbf{k}) \right]
\equiv \sum_{\chi=\pm} A^{(\chi)}(\omega;\mathbf{k}) {\cal P}_{5}^{(\chi)} ,
\label{spectral-A}
\end{equation}
and in the case under consideration equals
\begin{eqnarray}
A^{(\chi)}(\omega;\mathbf{k}) &=& ie^{-k_\perp^2 l^{2}}\sum_{\lambda=\pm} \sum_{n=0}^{\infty} 
 \frac{(-1)^n}{E_n^{(\chi)}}\Bigg\{
 \left[E_n^{(\chi)} \gamma^{0} 
 -\lambda v_F (k_3-\chi b)\gamma^3\right]\left[{\cal P}_{-}L_n\left(2 k_\perp^2 l^{2}\right)
-{\cal P}_{+}L_{n-1}\left(2 k_\perp^2 l^{2}\right)\right] \nonumber\\
&& + 2 \lambda v_F(\mathbf{k}_\perp\cdot\bm{\gamma}_\perp) L_{n-1}^1\left(2 k_\perp^2 l^{2}\right)
 \Bigg\}
\delta\left(\omega -\lambda  E_n^{(\chi)} \right).
\label{sp-funct-delta}
\end{eqnarray}
In the calculation of conductivities, we have to take into account that quasiparticles have a nonzero 
decay width (or equivalently, a finite scattering time). In order to model the corresponding effects, 
we replace the $\delta$ function in the spectral function (\ref{sp-funct-delta}) by a Lorentzian function, 
i.e.,
\begin{equation}
\delta(\omega -\lambda E_n^{(\chi)}) \to \frac{1}{\pi}\frac{\Gamma_n}{(\omega -\lambda E_n^{(\chi)})^2+\Gamma^2_n}.
\label{Lorentzian}
\end{equation}
Thus, we obtain
\begin{eqnarray}
A^{(\chi)}(\omega;\mathbf{k}) &=& \frac{ie^{-k_\perp^2 l^{2}}}{\pi}\sum_{\lambda=\pm}\sum_{n=0}^{\infty} 
 \frac{(-1)^n}{E_n^{(\chi)}}\Bigg\{
 \left[E_n^{(\chi)} \gamma^{0} 
 -\lambda v_F (k_3-\chi b)\gamma^3\right]\left[{\cal P}_{-}L_n\left(2 k_\perp^2 l^{2}\right)
-{\cal P}_{+}L_{n-1}\left(2 k_\perp^2 l^{2}\right)\right] \nonumber\\
&& + 2\lambda v_F(\mathbf{k}_\perp\cdot\bm{\gamma}_\perp) L_{n-1}^1\left(2 k_\perp^2 l^{2}\right)
 \Bigg\}\frac{\Gamma_n}{\left(\omega -\lambda E_n^{(\chi)} \right)^2+\Gamma_n^2}.
 \label{sp-funct-Gamma}
\end{eqnarray}
As is clear, the phenomenological modeling of the spectral functions is not complete 
without the calculation of the decay widths of quasiparticles in all Landau levels. The corresponding 
calculation of $\Gamma_n$ due to disorder/interaction will be very important for quantitative studies.
That, however, is beyond the scope of this study, which aims at revealing the qualitative
features of the magneto-transport characteristics in Weyl and Dirac semimetals. We expect, 
however, that the decay width in the LLL should be smaller than (or at most the same as) 
the decay widths in higher Landau levels. As soon as this assumption holds, our qualitative 
results should remain valid, i.e., the negative longitudinal magnetoresistivity will be 
realized in both Weyl and Dirac semimetals.

\section{Kubo formula}
\label{sec:KuboFormula}

According to the Kubo linear response theory, the direct current conductivity tensor
\begin{equation}
\sigma_{ij}  = \lim_{\Omega\to 0}\frac{\mbox{Im}\,\Pi_{ij}(\Omega+i0 ;\mathbf{0})}{\Omega}
\label{sigma-ij}
\end{equation}
is expressed through the Fourier transform of the current-current correlation function
\begin{equation}
\Pi_{ij}(\Omega;\mathbf{0}) =  e^2 v_F^2 T \sum_{k=-\infty}^{\infty} 
\int \frac{d^3 \mathbf{p}}{(2\pi)^3} \mbox{tr} \Big[ \gamma^i \bar{G}(i\omega_k ;\mathbf{p})  
\gamma^j \bar{G}(i\omega_k-\Omega; \mathbf{p})\Big].
\label{Pi_Omega_k}
\end{equation}
Note that this function is given in terms of the translation invariant part of the quasiparticle Green's function.
By making use of the spectral representation for the Green's function
\begin{equation}
\bar{G}(i\omega_k ;\mathbf{p})  = \int_{-\infty}^{\infty} \frac{d\omega A (\omega ;\mathbf{p}) }{i\omega_k+\mu-\omega},
\label{spectral-fun}
\end{equation}
we obtain the following standard representation for the current-current correlation function:
\begin{equation}
\Pi_{ij}(\Omega+i 0;\mathbf{0}) =
e^2 v_F^2  \int  d\omega \int d \omega^\prime 
\frac{n_F(\omega)-n_F(\omega^\prime)}{\omega -\omega^\prime-\Omega-i 0}
\int \frac{d^3 \mathbf{k}}{(2\pi)^3} 
\mbox{tr} \left[ \gamma^i A(\omega ;\mathbf{k}) \gamma^j A(\omega^\prime; \mathbf{k})\right], 
\label{Pi_ij}
\end{equation}
where $n_F(\omega)= 1/\left[e^{(\omega-\mu)/T}+1\right]$ is the Fermi distribution function. 

In the expression for the diagonal components of the current-current correlation function (\ref{Pi_ij}), 
the traces in the integrand are real [see Eqs.~(\ref{X11}) and (\ref{X33}) in Appendix~\ref{AppA}]. 
Therefore, in order to extract the imaginary part of $\Pi_{ii}(\Omega+i 0;\mathbf{0})$, we can use 
the identity
\begin{equation}
\frac{1}{\omega -\omega^\prime-\Omega-i 0} = {\cal P}\frac{1}{\omega -\omega^\prime-\Omega}
+i \pi \delta\left(\omega -\omega^\prime-\Omega\right) .
\label{prin-value}
\end{equation}
Taking this into account in Eq.~(\ref{Pi_ij}) and using the definition in Eq.~(\ref{sigma-ij}),
we derive a much simpler and more convenient expression for the diagonal components of the conductivity 
tensor:
\begin{eqnarray}
\sigma_{ii} &=& -\pi e^2 v_F^2  \sum_{\chi=\pm} \int \frac{d \omega }{4T\cosh^2\frac{\omega-\mu}{2T}}
\int \frac{d^3 \mathbf{k}}{(2\pi)^3} 
\mbox{tr} \left[ \gamma^i A^{(\chi)}(\omega ;\mathbf{k}) \gamma^i A^{(\chi)}(\omega; \mathbf{k})
 {\cal P}_{5}^{(\chi)}\right].
\label{sigma-ii-AA-chi}
\end{eqnarray}
(Here there is no sum over index $i$.)

The calculation of the off-diagonal components of the transverse conductivity $\sigma_{12}=-\sigma_{21}$ 
is complicated by the fact that the corresponding traces in Eq.~(\ref{Pi_ij}) are imaginary [see Eq.~(\ref{X12}) 
in Appendix~\ref{AppA}]. In this case, it is convenient to rewrite the expression for the current-current correlation 
function as follows:
\begin{eqnarray}
\Pi_{ij}(\Omega+i 0;\mathbf{0}) 
&=& e^2 v_F^2    \sum_{\chi=\pm} \int \frac{d^3 \mathbf{k}}{(2\pi)^3}  \int  d\omega n_F(\omega)
\mbox{tr} \Big[ \gamma^i A^{(\chi)}(\omega ;\mathbf{k}) \gamma^j \bar{G}^{(\chi)}_{\mu=0}(\omega-\Omega-i0; \mathbf{k})
 {\cal P}_{5}^{(\chi)}\nonumber\\
&&
 +\gamma^i \bar{G}^{(\chi)}_{\mu=0}(\omega +\Omega+i0 ;\mathbf{k}) \gamma^j A^{(\chi)}(\omega; \mathbf{k}) {\cal P}_{5}^{(\chi)}\Big],
\label{correlation-function}
\end{eqnarray}
where we used Eq.~(\ref{spectral-fun}) at $\mu=0$ in order to eliminate one of the energy integrations.
By substituting this result into 
Eq.~(\ref{sigma-ij}) and taking the limit $\Omega\to0$, we obtain
\begin{eqnarray}
\sigma_{ij} &=& e^2 v_F^2   \sum_{\chi=\pm} \mbox{Im} 
\int \frac{d^3 \mathbf{k}}{(2\pi)^3} \int  d\omega  n_F(\omega)
\mbox{tr} \Bigg[ 
\gamma^i  \frac{d\bar{G}^{(\chi)}_{\mu=0}(\omega+i0 ;\mathbf{k})}{d\omega} \gamma^j A^{(\chi)}(\omega;\mathbf{k}) 
 {\cal P}_{5}^{(\chi)}\nonumber\\
&&
 -\gamma^i A^{(\chi)}(\omega;\mathbf{k}) \gamma^j \frac{d\bar{G}^{(\chi)}_{\mu=0}(\omega-i0 ;\mathbf{k})}{d\omega} 
 {\cal P}_{5}^{(\chi)} \Bigg].
 \label{sigma-ij-AA-chi}
\end{eqnarray}
In principle this is valid for both the diagonal and off-diagonal components. In the case of the diagonal components, 
however, this is equivalent to the much simpler expression in Eq.~(\ref{sigma-ii-AA-chi}). In order to show their 
equivalency explicitly, one needs to integrate the expression in Eq.~(\ref{sigma-ij-AA-chi}) by parts and use the 
definition for the spectral function in Eq.~(\ref{spectral-A}). In the calculation of the off-diagonal components 
$\sigma_{ij}$, only the representation in Eq.~(\ref{sigma-ij-AA-chi}) is valid. 

Before concluding this section, it may be appropriate to mention that our analysis of the
conductivity in Dirac/Weyl semimetals in the presence of the magnetic field does not take into 
account the effect of weak localization/antilocalization \cite{anti-local1,anti-local2}. (For a recent 
study of weak localization and antilocalization in 3D Dirac semimetals, see Ref.~\cite{anti-local3}.) 
The corresponding quantum interference effects play an important role in weak magnetic fields and can 
even change the qualitative dependence of the conductivity/resistivity on the magnetic field. This expectation 
is also supported by the analysis of the experimental results \cite{Kim1307.6990}, where the 
signs of weak antilocalization are observed in weak magnetic fields. While the physics behind 
this effect is very interesting, it is not of prime interest for the purposes of our study here. 
Indeed, in the case of moderately strong magnetic fields considered, the effect of the weak
antilocalization is not expected to modify the qualitative behavior of the magnetoresistance.

\section{Longitudinal conductivity}
\label{sec:longitudinal}

As we discussed in the Introduction, the longitudinal conductivity is of special 
interest in Weyl semimetals because, as first suggested in Ref.~\cite{Nielsen}, 
it may reveal a unique behavior characteristic for these materials. Using Eq.~(\ref{sigma-ii-AA-chi}), we
find that the longitudinal conductivity is given by 
\begin{eqnarray}
\sigma_{33} &=&
\frac{e^2 v_F^2 }{2^4\pi^3 l^2 T}
\sum_{\chi} \sum_{n=0}^{\infty}
\int \frac{d \omega d k_3 }{\cosh^2\frac{\omega-\mu}{2T}}
\frac{\Gamma_n^2\left[\left(\omega-s_\perp\chi v_F(k_3-\chi b)\right)^2+2n\epsilon_{L}^2+\Gamma_n^2\right]^2}
{\left[\left(\omega -E_n^{(\chi)} \right)^2+\Gamma_n^2\right]^2
\left[ \left(\omega +E_{n}^{(\chi)} \right)^2+\Gamma_{n}^2\right]^2} \nonumber \\
&+&  \frac{e^2 v_F^2 }{2^4\pi^3 l^2 T}
\sum_{\chi} \sum_{n=1}^{\infty}
\int \frac{d \omega d k_3 }{\cosh^2\frac{\omega-\mu}{2T}}
\frac{\Gamma_n^2\left[\left(\omega+s_\perp\chi v_F(k_3-\chi b)\right)^2+2n\epsilon_{L}^2+\Gamma_n^2\right]^2}
{\left[\left(\omega -E_n^{(\chi)} \right)^2+\Gamma_n^2\right]^2
\left[ \left(\omega +E_{n}^{(\chi)} \right)^2+\Gamma_{n}^2\right]^2}\nonumber \\
&-&  \frac{e^2 v_F^2 }{\pi^3 l^2 T}
\sum_{\chi}\sum_{n=1}^{\infty}
\int \frac{d \omega d k_3 }{\cosh^2\frac{\omega-\mu}{2T}}
\frac{\Gamma_n^2 \omega^2  n\epsilon_{L}^2}
{\left[\left(\omega -E_n^{(\chi)} \right)^2+\Gamma_n^2\right]^2
\left[ \left(\omega +E_{n}^{(\chi)} \right)^2+\Gamma_{n}^2\right]^2},
\label{conductivity-longitudinal}
\end{eqnarray}
where $\epsilon_{L} \equiv v_F/l \equiv v_F\sqrt{|eB|/c}$ is the Landau energy scale. 

Before analyzing the complete expression, it is instructive to extract the LLL contribution
 $\sigma_{33}^{\rm (LLL)}$ to the  longitudinal conductivity. It is given by the following exact 
result:
\begin{eqnarray}
\sigma_{33}^{\rm (LLL)} &=&\frac{e^2 v_F^2 }{2^4\pi^3 l^2 T}
\sum_{\chi} 
\int \frac{d \omega d k_3 }{\cosh^2\frac{\omega-\mu}{2T}}
\frac{\Gamma_0^2}{\left[\left(\omega+s_\perp\chi v_F(k_3-\chi b)\right)^2+\Gamma_0^2\right]^2} 
=\frac{e^2 v_F}{4 \pi^2 l^2 \Gamma_0}
=\frac{e^2 v_F |eB|}{4 \pi^2 c \Gamma_0}.
\label{sigma33LLL}
\end{eqnarray}
This is a {\it topological contribution} associated with the chiral anomaly, which is generated entirely 
on the LLL in the presence of a magnetic field \cite{Ambjorn}. It is completely independent of the 
temperature and the chemical potential. This result agrees also with the corresponding result obtained by 
using the semiclassical Boltzmann kinetic equation in Refs.~\cite{Nielsen,Aji,Son}. By comparing the
expression in Eq.~(\ref{sigma33LLL}) with those in Refs.~\cite{Nielsen,Aji,Son}, we see that the 
quasiparticle width $\Gamma_0$ is related to the collision time as follows: $\Gamma_0=\hbar/\tau$.

It is interesting that the origin of the topological contribution in Eq.~(\ref{sigma33LLL}) is 
intimately connected with the spatial dimensional reduction $3 \to 1$ in the LLL dynamics \cite{reduction}. 
The dimensional reduction of the LLL states can be made explicit by noting that the 
propagator of the corresponding quasiparticles of given chirality $\chi$ (Weyl node), according to 
Eq.~(\ref{full-propagator}), is given by
\begin{equation}
\bar{G}^{(\chi)}_{LLL}(\omega,\mathbf{k})=ie^{-k^2_{\perp}l^2}\frac{(\omega+\mu)\gamma^0-v_F(k_3-\chi b)\gamma^3}
{(\omega+\mu)^2-v^2_F(k_3-\chi b)^2}\,(1-is_{\perp}\gamma^1\gamma^2).
\label{LLLprop}
\end{equation}
This propagator implies that the LLL modes are characterized by a one-dimensional form of the 
relativistic-like dispersion relation $\omega^{(\chi)}=-\mu \pm v_F(k_3-\chi b)$, which is independent 
of the magnetic field. The final expression for the topological contribution is proportional to the magnetic 
field only because the LLL density of states is determined by the strength of the field.

The remaining higher Landau level (HLL) contribution to the  longitudinal conductivity is given 
by the following expression:
\begin{eqnarray}
\sigma_{33}^{\rm (HLL)} &=&\frac{e^2 v_F^2 }{4\pi^3 l^2 T}
\sum_{n=1}^{\infty} 
\int \frac{d \omega d k_3 }{\cosh^2\frac{\omega-\mu}{2T}}
\frac{\Gamma_n^2\left[
\left(\omega^2+E_n^2+\Gamma_n^2\right)^2- 4n \epsilon_{L}^2\omega^2 \right]}
{\left[\left(\omega -E_n  \right)^2+\Gamma_n^2\right]^2
\left[ \left(\omega +E_{n}  \right)^2+\Gamma_{n}^2\right]^2},
\label{sigma33HLL}
\end{eqnarray}
where $E_n=v_F \sqrt{k_3^2 + 2 n |eB|/c}$. Note that the integration over $k_3$ in the last 
expression can be performed analytically. Moreover, in the limit of zero temperature, the remaining 
integration over $\omega$ can be performed as well. The corresponding explicit results are presented 
in Eqs.~(\ref{sigma33HLL_K3}) and (\ref{sigma33HLL_omega}) in Appendix~\ref{AppB}.

\begin{figure}
\begin{center}
\includegraphics[width=.45\textwidth]{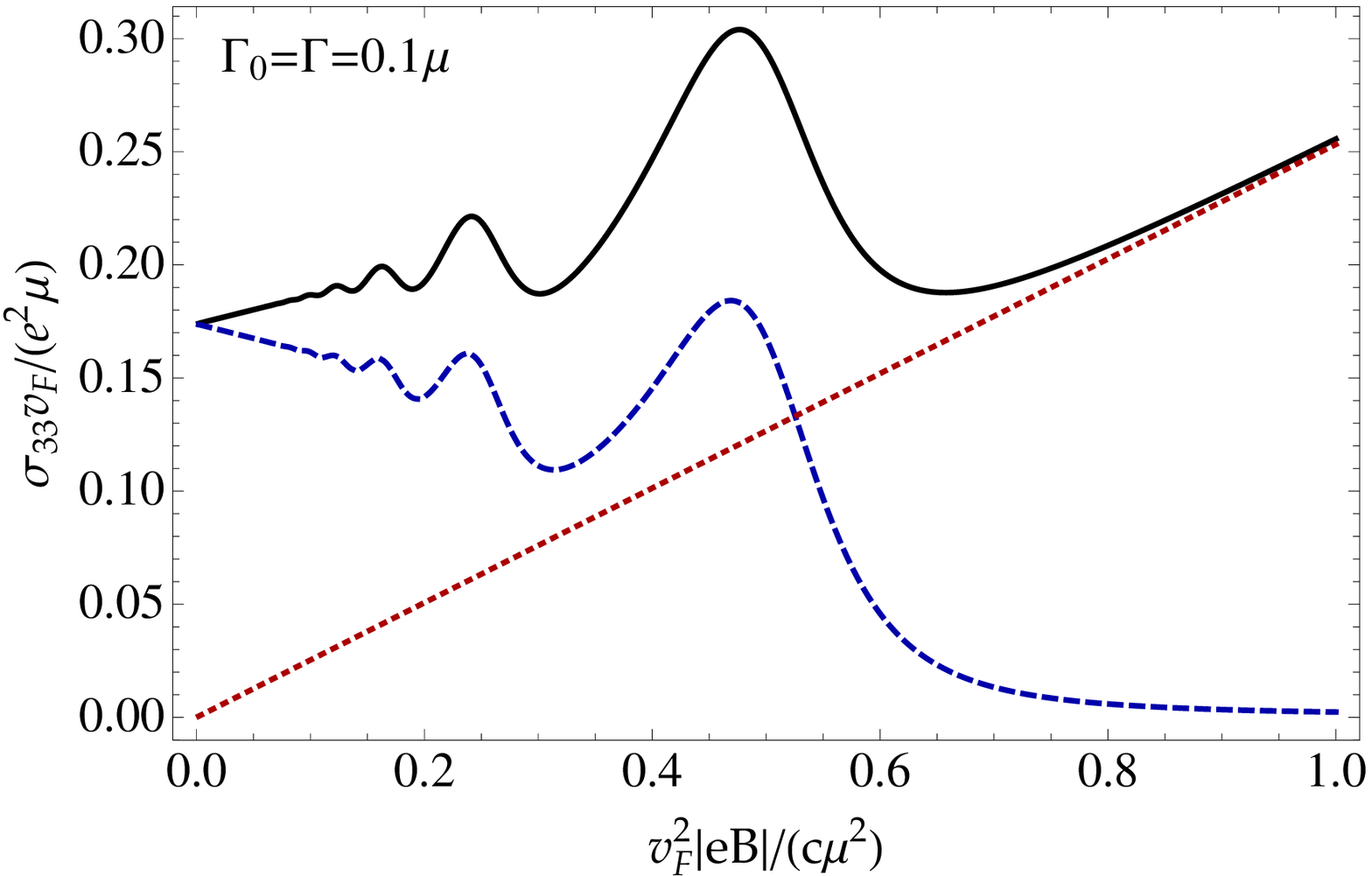}
\hspace{5mm}
\includegraphics[width=.45\textwidth]{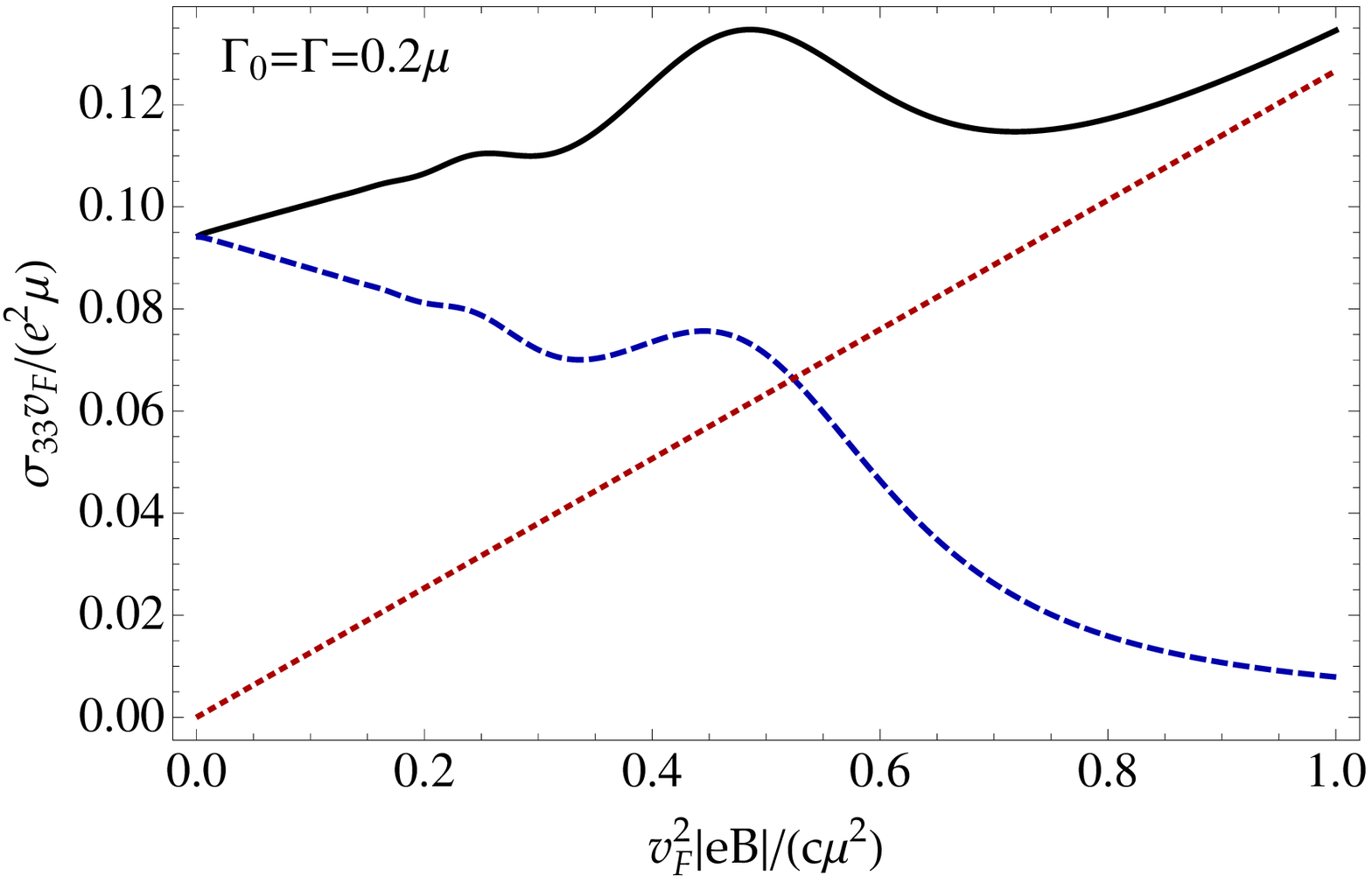}\\
\includegraphics[width=.45\textwidth]{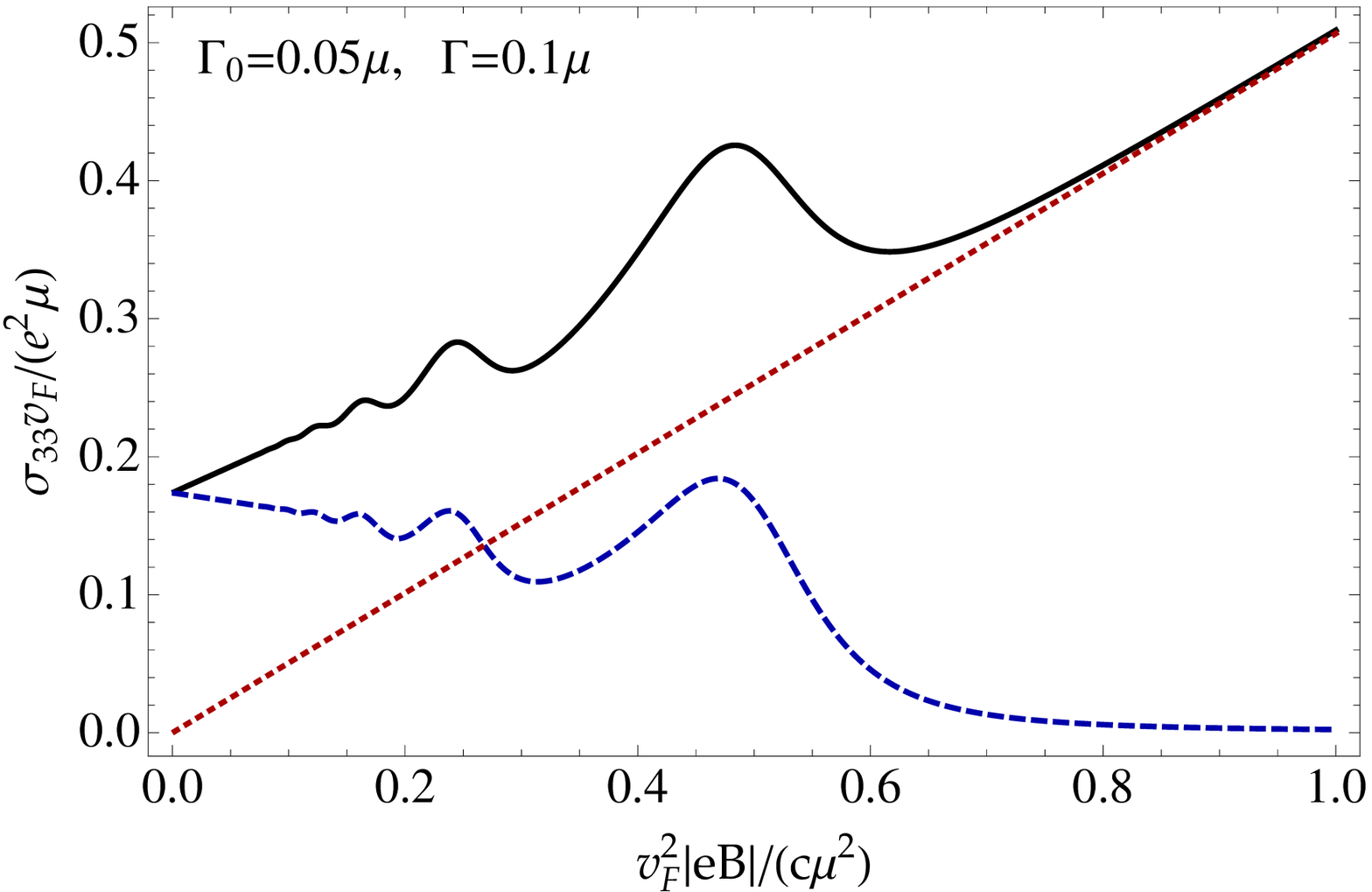}
\hspace{5mm}
\includegraphics[width=.45\textwidth]{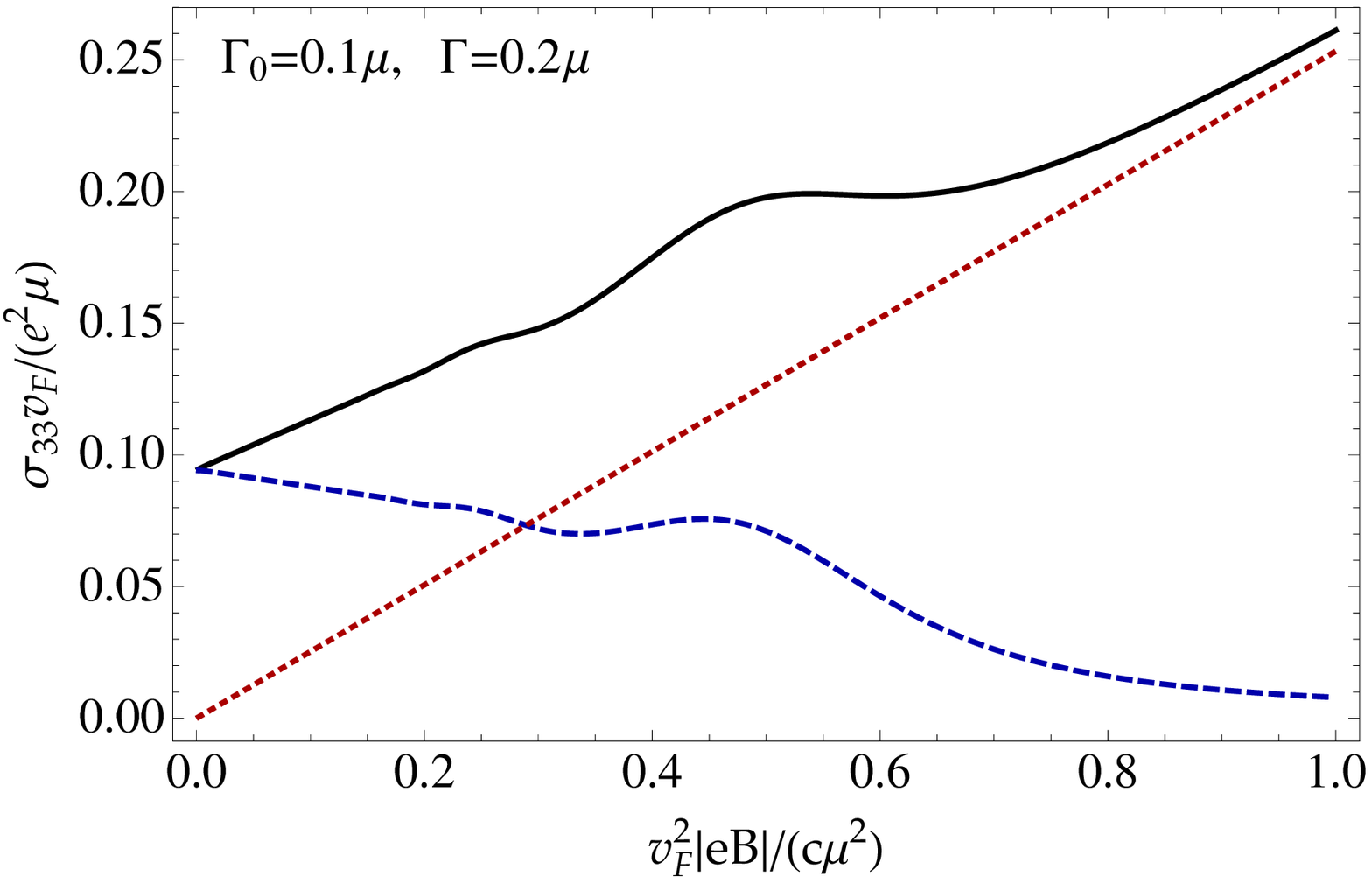}
\caption{(Color online) Longitudinal conductivity $\sigma_{33}$ at zero temperature as a function of the magnetic field. 
The solid line shows the complete result, the dashed line shows the contribution without the lowest 
Landau level, and the dotted line shows the topological contribution of the lowest Landau level alone.
The quasiparticle width in higher Landau levels is $\Gamma=0.1\mu$ (left panels) and $\Gamma=0.2\mu$ (right panels).
The LLL quasiparticle width is the same (upper panels) or half (lower panels) the width in higher Landau levels.}
\label{fig:Fermi}
\end{center}
\end{figure}

The numerical results for the longitudinal magnetoconductivity as functions of $v^2_F|eB|/\mu^2c$ are 
plotted in Fig.~\ref{fig:Fermi} for two fixed values of the quasiparticle widths in the higher Landau 
levels, i.e., $\Gamma=0.1\mu$ (left panels) and $\Gamma=0.2\mu$ (right panels), and with the two 
possible choices of the LLL quasiparticle width $\Gamma_0$, i.e., the same (upper panels) or two 
times smaller (lower panels) than the width in the higher Landau levels. The LLL contribution is 
shown by the red dotted line, the HLL contribution is shown by the blue dashed line, and the 
complete expression for the longitudinal magnetoconductivity, 
$\sigma_{33}=\sigma_{33}^{\rm (LLL)}+\sigma_{33}^{\rm (HLL)}$, is shown by the black solid line.
Leaving aside the characteristic Shubnikov-de Haas oscillations, we see that the HLL contribution has an 
overall tendency to decrease with increasing the field. In spite of that, the total longitudinal magnetoconductivity, 
which also includes the linearly increasing topological LLL contribution, has the opposite tendency.

\begin{figure}
\begin{center}
\includegraphics[width=.45\textwidth]{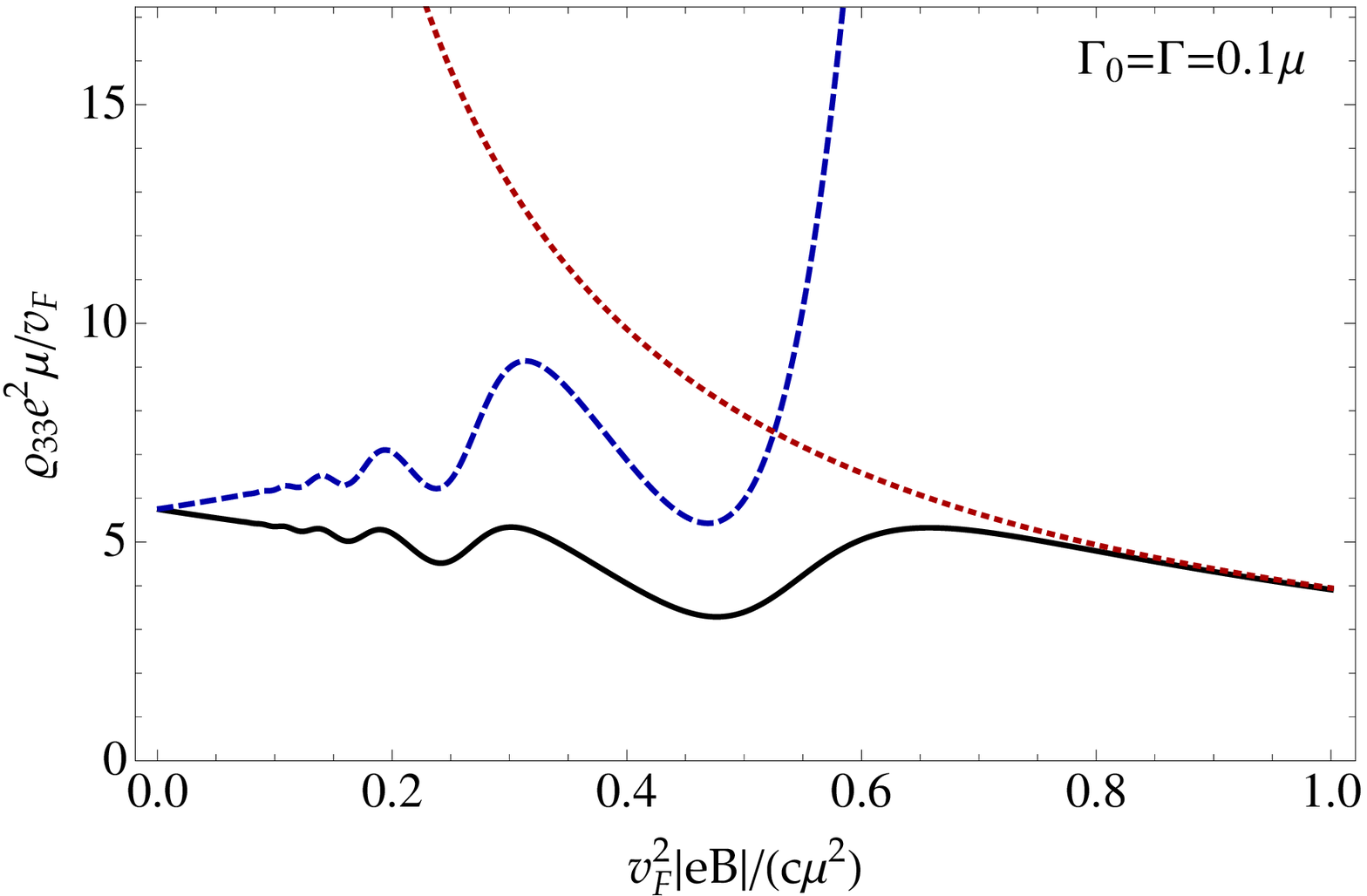}
\hspace{5mm}
\includegraphics[width=.45\textwidth]{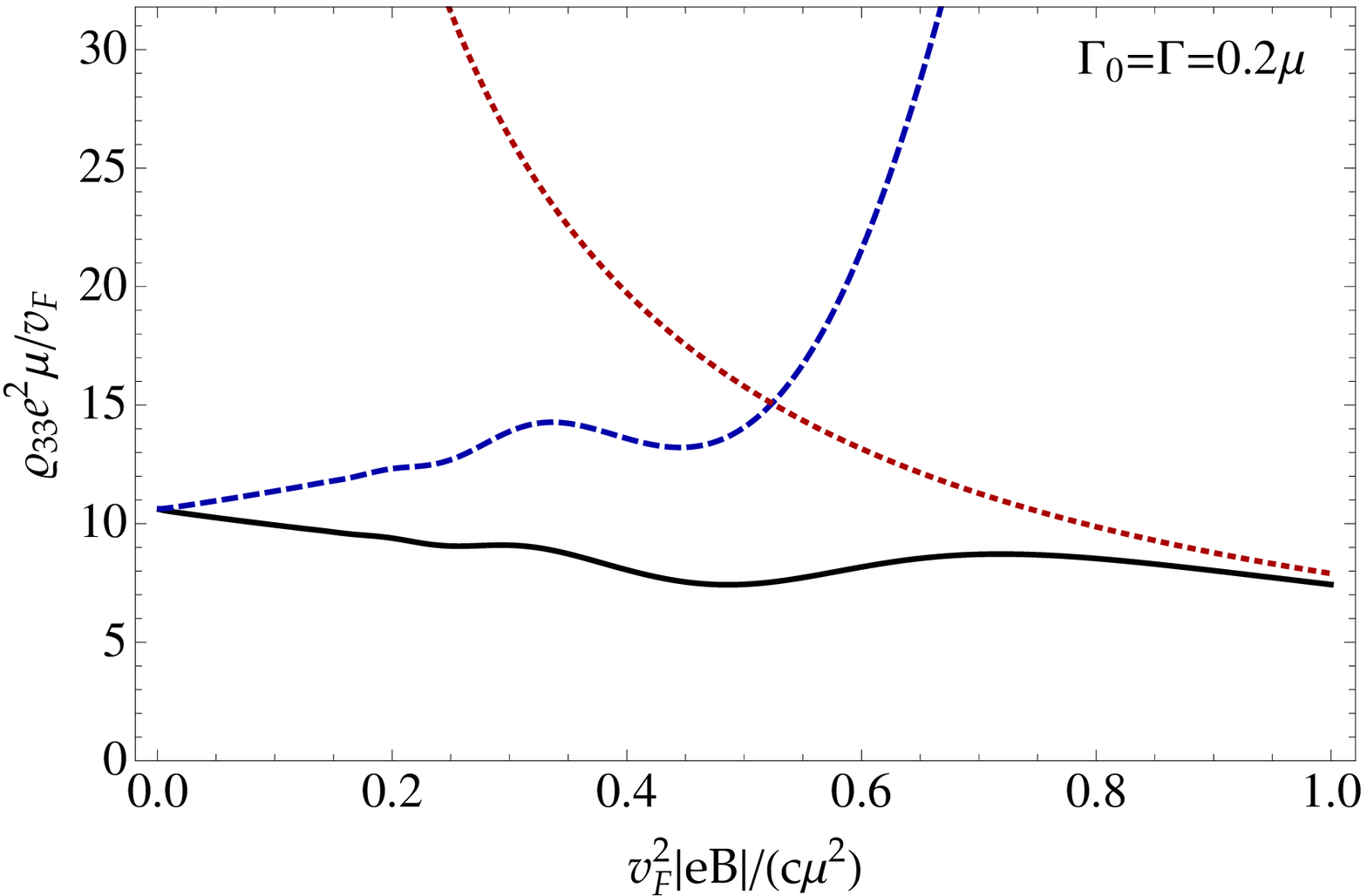}\\
\includegraphics[width=.45\textwidth]{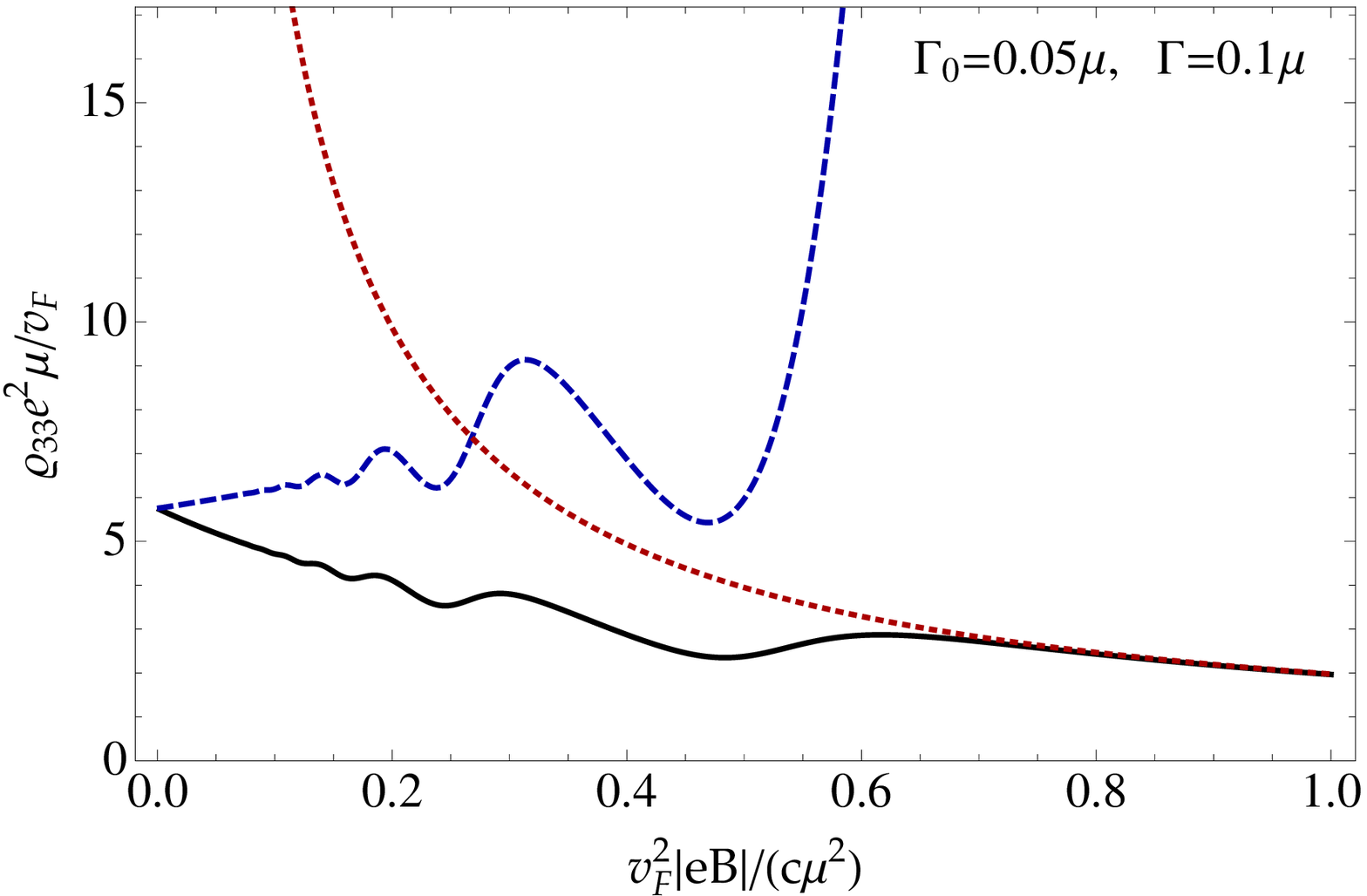}
\hspace{5mm}
\includegraphics[width=.45\textwidth]{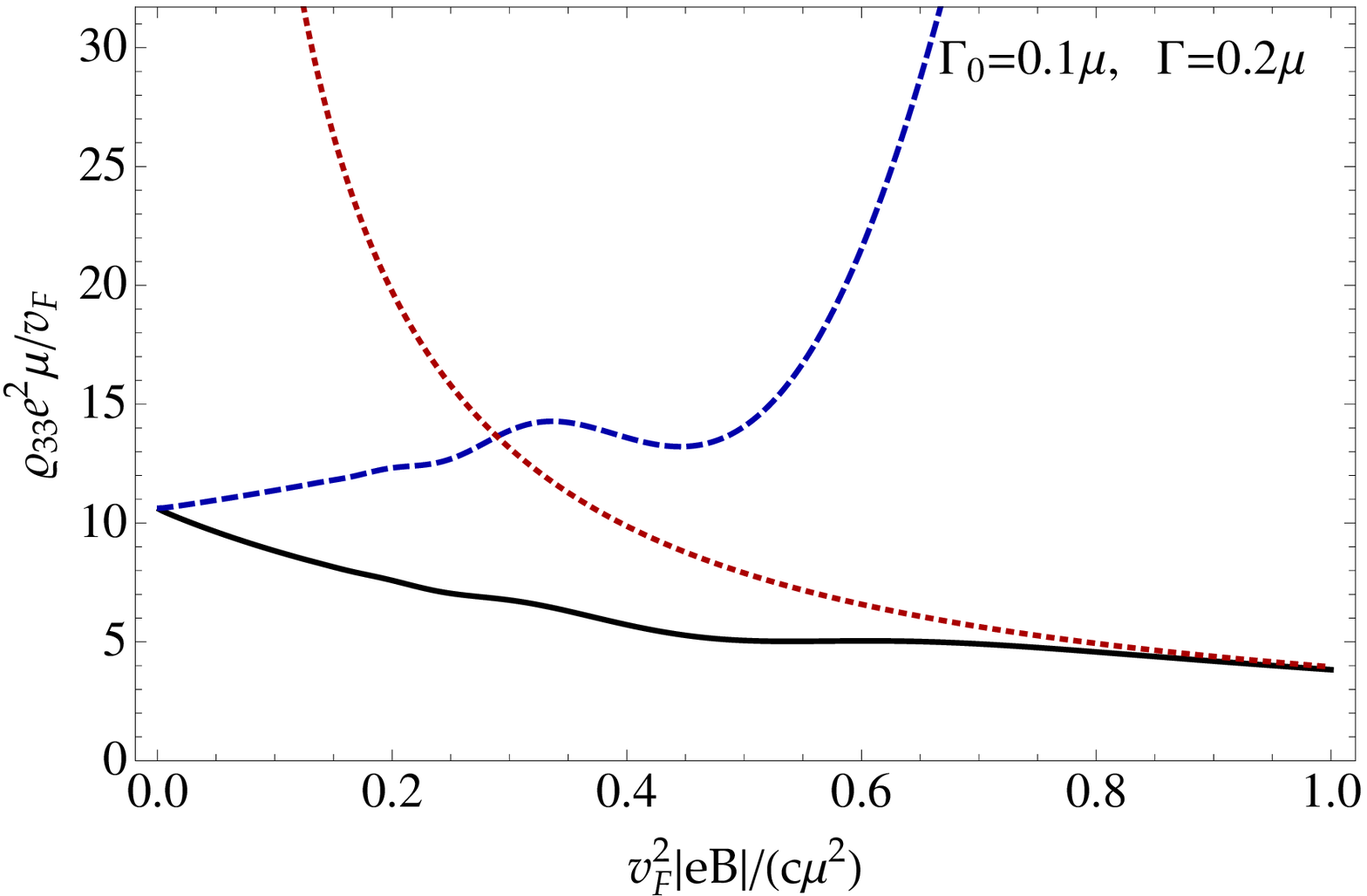}
\caption{(Color online) Longitudinal resistivity $\rho_{33}$ at zero temperature as a function of the magnetic field. 
The solid line shows the complete result, the dashed line shows the contribution without the lowest 
Landau level, and the dotted line shows the topological contribution of the lowest Landau level alone.
The quasiparticle width is $\Gamma=0.1\mu$ (left panels) and $\Gamma=0.2\mu$ (right panels).
The LLL quasiparticle width is the same (upper panels) or half (lower panels) the width 
in higher Landau levels.}
\label{fig:longitudinal}
\end{center}
\end{figure}

Taking into account that $\sigma_{13}=\sigma_{31}=\sigma_{23}=\sigma_{32}=0$ and 
using $\sigma_{33}$ calculated above, we also find the longitudinal magnetoresistivity. 
It is given by $\rho_{33}=1/\sigma_{33}$. The corresponding numerical results are plotted in 
Fig.~\ref{fig:longitudinal} as functions of $v^2_F|eB|/\mu^2c$. Oscillations of 
magnetoresistivity connected with the Shubnikov-de Haas effect are clearly seen 
in the left panels in Fig.~\ref{fig:longitudinal}, which show the results for a smaller value 
of the quasiparticle width $\Gamma=0.1 \mu$ in higher Landau levels. The 
oscillations in the case of twice as large width, $\Gamma=0.2 \mu$, are not as well 
pronounced. The longitudinal magnetoresistivity in the case with the LLL
quasiparticle width two times smaller than the width of higher Landau levels is plotted in 
the two lower panels. Overall, the longitudinal 
magnetoresistivity decreases as the magnetic field grows. As we mentioned in the Introduction, 
this phenomenon is known in the literature as negative magnetoresistivity. As is clear from our results 
in Fig.~\ref{fig:longitudinal}, the negative longitudinal magnetoresistivity is exclusively due to 
the LLL contribution\cite{Nielsen} which in turn is connected with the chiral anomaly \cite{anomaly}.

We would like to emphasize that we did not assume in our calculations that $\Gamma_0$ 
is much less than the quasiparticle width in higher Landau levels. This assumption was made in 
semiclassical calculations in Refs.~\cite{Nielsen,Aji,Son} due to the fact that the 
quasiparticle width $\Gamma_0$ in the LLL is not equal to zero only because of the internode 
scatterings. This is unlike the quasiparticle width in higher Landau levels where intranode scatterings 
contribute too. Since Weyl nodes are separated by the distance $2b$ in momentum 
space in Weyl semimetals, internode scattering processes are less efficient compared to intranode 
ones. Therefore, it is usually assumed that $\Gamma_0$ is much less than $\Gamma_n$ in higher 
Landau levels $n \ge 1$. Although we did not make this assumption, we still observe the negative 
longitudinal magnetoresistivity. It is also important to emphasize another point. After the change 
of the integration variable $k_3  \to  k_{\rm new}^3\equiv k_3-\chi b$, the chiral shift $b$ does not 
enter in the longitudinal magnetoconductivity (\ref{conductivity-longitudinal}) and affects the result 
only indirectly through the quasiparticle width \cite{Nielsen}. Since our results show that the 
negative longitudinal magnetoresistivity takes place even when the LLL quasiparticle width 
$\Gamma_0$ is comparable to the width $\Gamma_n$ in the higher Landau levels, we conclude that 
this phenomenon is quite robust and will also take place in Dirac semimetals as well.

\section{Transverse conductivity}
\label{sec:transverse}

\subsection{Diagonal components of the transverse conductivity}

In this subsection, we calculate the diagonal component $\sigma_{11}=\sigma_{22}$
of the transverse conductivity by starting from the definition in Eq.~(\ref{sigma-ii-AA-chi}).
The key intermediate steps of the derivation are given in Appendix~\ref{AppB}. The final 
result takes the following form:
\begin{eqnarray}
\sigma_{11} &=&   \frac{e^2 v_F^2 }{4\pi^3 l^2 T}
\sum_{n=0}^{\infty}
\int \frac{d \omega d k_3 }{\cosh^2\frac{\omega-\mu}{2T}}
 \frac{\Gamma_{n+1} \Gamma_{n}
 \left[ \left(\omega^2+E_{n}^2+\Gamma_{n}^2 \right) \left(\omega^2+E_{n+1}^2+\Gamma_{n+1}^2 \right)-4(v_F k_3)^2\omega^2\right]}
{\left[\left(E_{n}^2+\Gamma_{n}^2 -\omega^2\right)^2+4\omega^2 \Gamma_{n}^2 \right]
\left[\left(E_{n+1}^2+\Gamma_{n+1}^2 -\omega^2\right)^2+4\omega^2 \Gamma_{n+1}^2 \right]}.  
\label{sigma11_text}
\end{eqnarray}
In the limit of zero temperature, we can easily integrate over $\omega$ and $k_3$. The corresponding 
analytical result is presented in Eq.~(\ref{sigma11}) in Appendix~\ref{AppB}.

\begin{figure}
\begin{center}
\includegraphics[width=.45\textwidth]{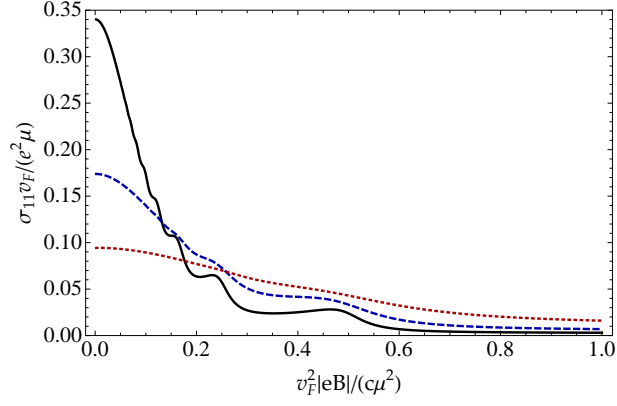}
\caption{(Color online) Diagonal components of the transverse conductivity $\sigma_{11}=\sigma_{22}$ 
at zero temperature as a function of the magnetic field. The quasiparticle width is 
$\Gamma=0.05\mu$ (black solid line), $\Gamma=0.1\mu$ (blue dashed line), and 
$\Gamma=0.2\mu$ (red dotted line). The sum over Landau levels includes 
$n_{\rm max} =10^{4}$ levels.}
\label{fig:sigma11}
\end{center}
\end{figure}

The numerical results for the transverse diagonal conductivity $\sigma_{11}$ 
as a function of $v^2_F|eB|/(\mu^2c)$ are shown in Fig.~\ref{fig:sigma11} for three different values of the quasiparticle 
width. Just as in the case of longitudinal conductivity, the Shubnikov-de Haas oscillations are clearly seen for smaller 
values of the width, but gradually disappear when the width becomes larger. In all cases, however, the transverse 
diagonal conductivity has an overall tendency to decrease with increasing the field. 

\subsection{Off-diagonal components of the transverse conductivity}

In order to calculate the off-diagonal components of the transverse conductivity, we use Eq.~(\ref{sigma-ij-AA-chi}).
Let us start from the simplest case when $\Gamma_n\to 0$. In this limit, the spectral function (\ref{sp-funct-delta})
contains $\delta$ function and the analysis greatly simplifies. The corresponding result reads as
\begin{eqnarray}
\sigma_{12} &=& - \frac{e^2 v_F^2 s_\perp}{4\pi^2 l^2} 
\sum_{\lambda,\lambda^\prime=\pm}
\sum_{n}
\int d k_3   
\frac{n_F\left(\lambda^\prime E_{n}\right)-n_F\left(\lambda^\prime E_{n+1}\right)}{\left(E_{n} -\lambda E_{n+1}\right)^2}
\left(1- \frac{\lambda \left(v_F k_{3}\right)^2 }{ E_{n+1} E_{n}} \right)\nonumber\\
&&+\frac{e^2 v_F^2}{8\pi^2 l^2} \sum_{\chi=\pm} \sum_{\lambda,\lambda^\prime=\pm}
\sum_{n,n^\prime}
\int d k_3 \frac{n_F\left(\lambda E_{n}^{(\chi)} \right)}{ E_{n}^{(\chi)} E_{n^\prime}^{(\chi)}}\frac{\chi v_F(k_3-\chi b)}
{ \lambda^\prime E_{n}^{(\chi)} - \lambda E_{n^\prime}^{(\chi)} }\left(\delta_{n-1,n^\prime}+\delta_{n,n^\prime-1}\right)
\nonumber\\
&= &- \frac{e^2 s_\perp}{4\pi^2} 
\sum_{n}\alpha_{n}\int d k_3 \frac{ \sinh\frac{\mu}{T}}{\cosh\frac{E_n}{T}+\cosh\frac{\mu}{T}}
-\frac{e^2}{8\pi^2} \sum_{\chi=\pm}\chi\int dk_3\frac{\sinh\frac{v_F(k_3-\chi b)}{T}}{\cosh\frac{v_F(k_3-\chi b)}{T}+\cosh\frac{\mu}{T}},
\label{sigma12-T}
\end{eqnarray}
where $\alpha_{n}=2-\delta_{n,0}$ is the spin degeneracy of the Landau levels. The first 
term in the last line is associated with a nonzero density of charge carriers. It comes from the occupied 
Landau levels and, as expected, depends on the temperature, chemical potential, and magnetic 
field. In contrast, the last term in Eq.~(\ref{sigma12-T}) is a topological vacuum contribution (which 
is present even at $\mu=0$) and comes exclusively from the lowest Landau level. Such a contribution
is a specific feature of Weyl semimetals and is directly related to the 
anomalous Hall effect \cite{Haldane}, which is produced by the dynamical Chern-Simons term in 
Weyl semimetals \cite{Burkov1,Grushin,ZyuzinBurkov,Franz,Goswami}. This topological (anomalous) 
contribution is independent of the temperature, chemical potential, and magnetic field and equals
\begin{eqnarray}
\sigma_{12,{\rm anom}} &=& 
-\frac{e^2}{8\pi^2 v_F} T \left.
\ln\frac{\cosh\frac{v_F(k_3-b)}{T}+\cosh\frac{\mu}{T} }{\cosh\frac{v_F(k_3+ b)}{T}+\cosh\frac{\mu}{T} }
\right|_{k_3=-\infty}^{k_3=\infty}
 =\frac{e^2 b}{2\pi^2}.
\label{anomaly-contribution}
\end{eqnarray}
As usual in calculations of anomalous quantities, the integral form of the topological contribution 
in the last term in Eq.~(\ref{sigma12-T}) should be treated with care. Indeed, while 
separate left- and right-handed contributions appear to be poorly defined because of a 
linear divergency, the sum of both chiralities results in a convergent integral. 

It should be noted that there is no interference between the topological contribution and the 
remaining contribution due to the finite density of charge carriers. We should also emphasize that the anomalous
contribution (\ref{anomaly-contribution}) will be present even in Dirac semimetals in a magnetic 
field because, as we discussed in the Introduction, $\mathbf{b} \ne 0$ is generated in Dirac 
semimetals by the Zeeman interaction or dynamically \cite{engineering}. The anomalous contribution 
(\ref{anomaly-contribution}) unambiguously distinguishes a Weyl semimetal from a Dirac one only in the absence 
of a magnetic field. In such a case, nonzero $\mathbf{b}$ breaks time reversal symmetry in Weyl 
semimetals and provides finite $\sigma_{12}$ unlike the case of Dirac semimetals where $\mathbf{b}$ 
is absent and, therefore, time reversal symmetry is preserved and $\sigma_{12}$ vanishes. 

\begin{figure}
\begin{center}
\includegraphics[width=.45\textwidth]{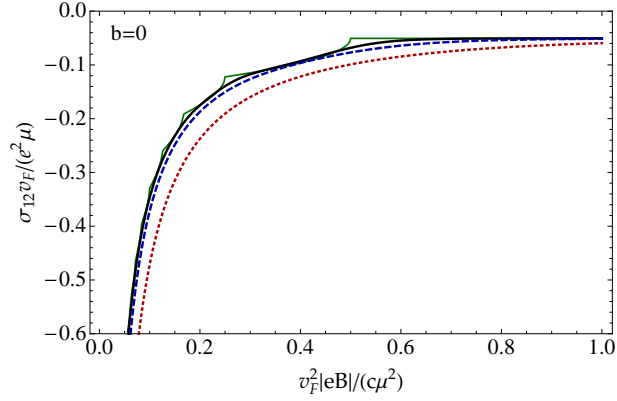}
\caption{(Color online) Off-diagonal components of the transverse conductivity $\sigma_{12}=-\sigma_{21}$ 
as a function of the magnetic field for the vanishing chiral shift, $\mathbf{b}=0$. The results are shown 
for $\Gamma = T=0$ (green thin solid line), $\Gamma \to T=0.05 \mu$ (black solid line), 
$\Gamma \to T=0.1\mu$ (blue dashed line), and $\Gamma \to T=0.2\mu$ (red dotted line). 
If $b \ne 0$, the conductivity will simply shift by $e^2 b/(2\pi^2)$.}
\label{fig:sigma12}
\end{center}
\end{figure}

In the limit of zero temperature, the complete expression for the off-diagonal conductivity is given by the 
following analytical expression:
\begin{eqnarray}
\sigma_{12} &=&\frac{e^2 b}{2\pi^2}
- \frac{e^2 s_\perp \mbox{sgn}(\mu)}{4\pi^2}  \sum_{n} \alpha_{n} \int dk_3  
 \theta\left(|\mu|-|E_{n}|\right)
 =\frac{e^2 b}{2\pi^2}
 - \frac{e^2 s_\perp \mbox{sgn}(\mu)}{2\pi^2 v_F}  \sum_{n=0}^{n_{\rm max}} \alpha_{n}\sqrt{\mu^2-2nv_F^2|eB|/c},
\label{sigma12}
\end{eqnarray}
where $n_{\rm max}$ is given by the integer part of $\mu^2/(2\epsilon_L^2)$ and has the meaning of
the Landau level index in the highest occupied Landau level. The off-diagonal component of the 
conductivity is plotted in Fig.~\ref{fig:sigma12} (green thin solid line).

It may be appropriate to note here that the expression for the off-diagonal component of the
conductivity in the case of quasiparticles with nonzero widths, modeled by the Lorentzian 
distribution (\ref{Lorentzian}), is not as convenient or even useful as the above expression. 
In fact, unlike the similar expressions for the diagonal components of the conductivity, 
off-diagonal component $\sigma_{12}$ contains a formally divergent sum over the Landau levels
when $\Gamma_n\neq 0$. This can be checked by first explicitly calculating the integrals 
over the energy and the longitudinal momentum, and then examining the contributions of 
the Landau levels with large values of Landau index $n$. The corresponding contributions 
are suppressed only as $1/\sqrt{n}$ when $n\to\infty$ and, therefore, cause a divergence in the 
sum. From the physics viewpoint, the origin of the problem is rooted in the use of the simplest Lorentzian 
model (\ref{Lorentzian}) for the quasiparticle spectral function with nonzero quasiparticle widths. 
The corresponding distribution falls off too slowly as a function of the energy. As a result, the 
Landau levels with very large $n$, which are completely empty and should not have much 
of an effect on the conductivity, appear to give small individual contributions (suppressed 
only as $1/\sqrt{n}$) that collectively cause a divergence.  

In order to illustrate the problem in the simplest possible mathematical form, we can mimic 
the result of the integration by the following approximate form:
\begin{eqnarray}
\sigma_{12} &\simeq & - \frac{e^2 s_\perp}{4\pi^3} \sum_{n}\alpha_n \int d k_3
\left[\arctan\frac{E_n+\mu}{\Gamma}-\arctan\frac{E_n-\mu}{\Gamma}\right]\nonumber\\
&= &- \frac{e^2 s_\perp}{\sqrt{2}  \pi^2 v_F}  \sum_{n}\alpha_n
\frac{\Gamma\mu}{\sqrt{2n\epsilon_{L}^2+\Gamma^2-\mu^2+\sqrt{(2n\epsilon_{L}^2+\Gamma^2-\mu^2)^2+4\Gamma^2\mu^2}}},
\end{eqnarray}
which correctly captures the zero quasiparticle width approximation on the one hand and 
shares the same problems as the exact result obtained from the expression in the model with
the Lorentzian quasiparticle widths. 

Ideally, in order to better incorporate the effects of finite widths of quasiparticles in the calculation
of the off-diagonal component of the conductivity, one has to use a better and more realistic model 
for the spectral function. Such a task is beyond the scope of this paper. An alternative sensible 
way to incorporate the effect of the finite widths of quasiparticles is suggested by the 
finite-temperature expression in Eq.~(\ref{sigma12-T}). It is not unreasonable at all to assume that 
a nonzero but small width $\Gamma$ may be mimicked by the effects of a small temperature 
$T\simeq \Gamma$. Then, by making use of the expression in Eq.~(\ref{sigma12-T}) with the
corresponding replacement, we can roughly estimate the effect of a small nonzero width. The 
corresponding numerical results for $\Gamma\to T=0.05\mu$, $\Gamma\to T=0.1\mu$, and 
$\Gamma\to T=0.2\mu$ are shown in Fig.~\ref{fig:sigma12} as the solid black line, the blue 
dashed line, and the red dotted line, respectively.

\begin{figure}
\begin{center}
\includegraphics[width=.45\textwidth]{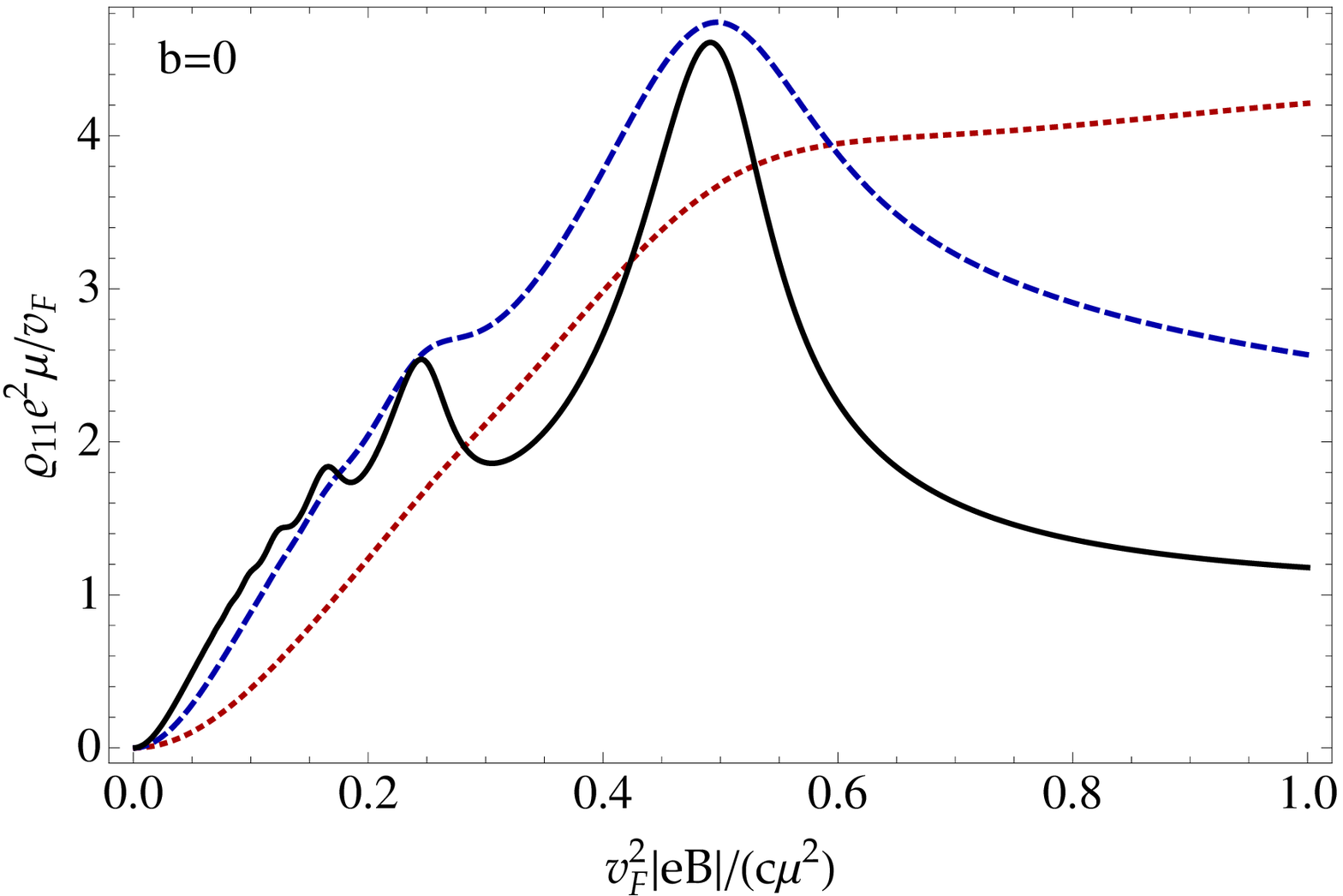}
\hspace{5mm}
\includegraphics[width=.45\textwidth]{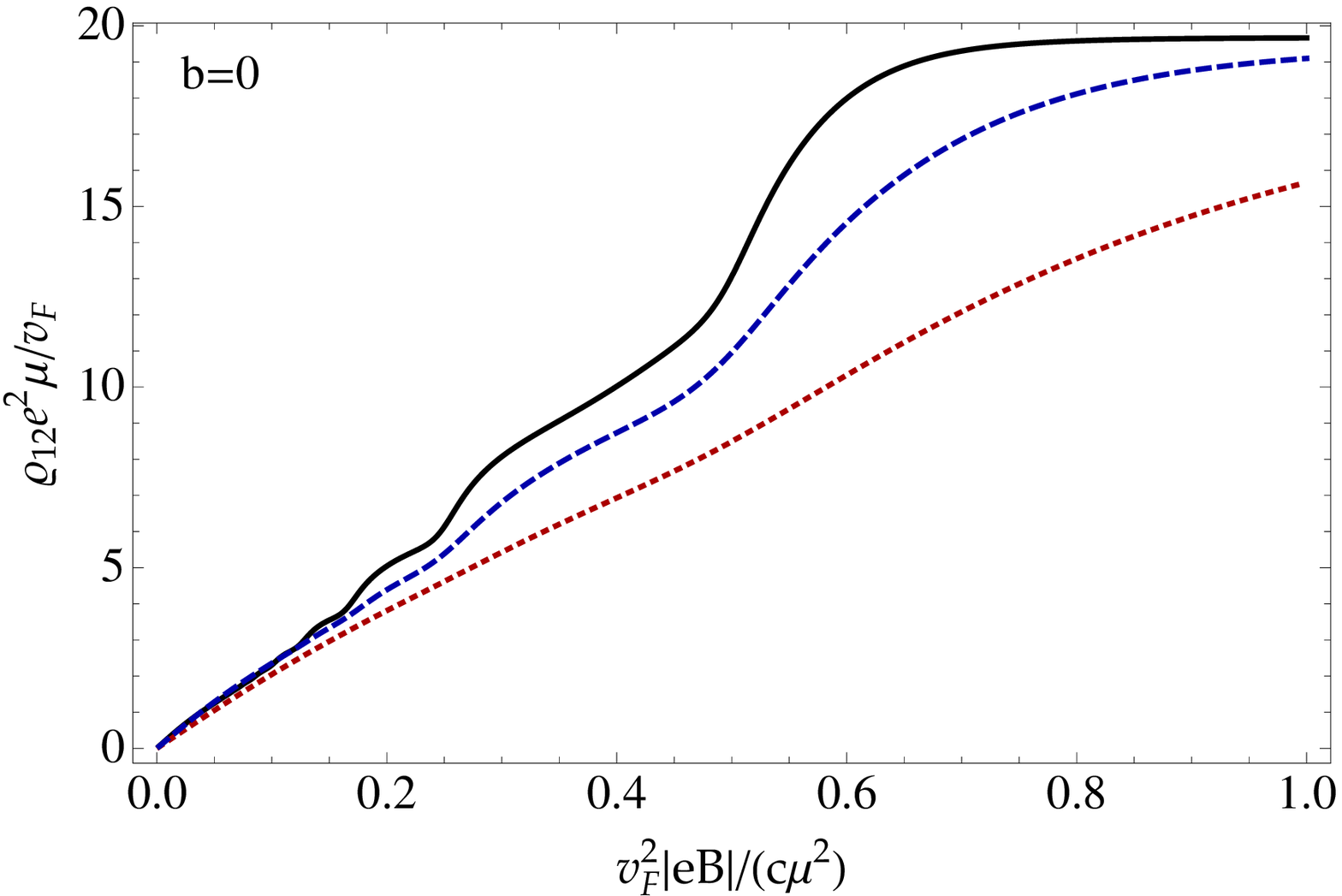}\\
\includegraphics[width=.45\textwidth]{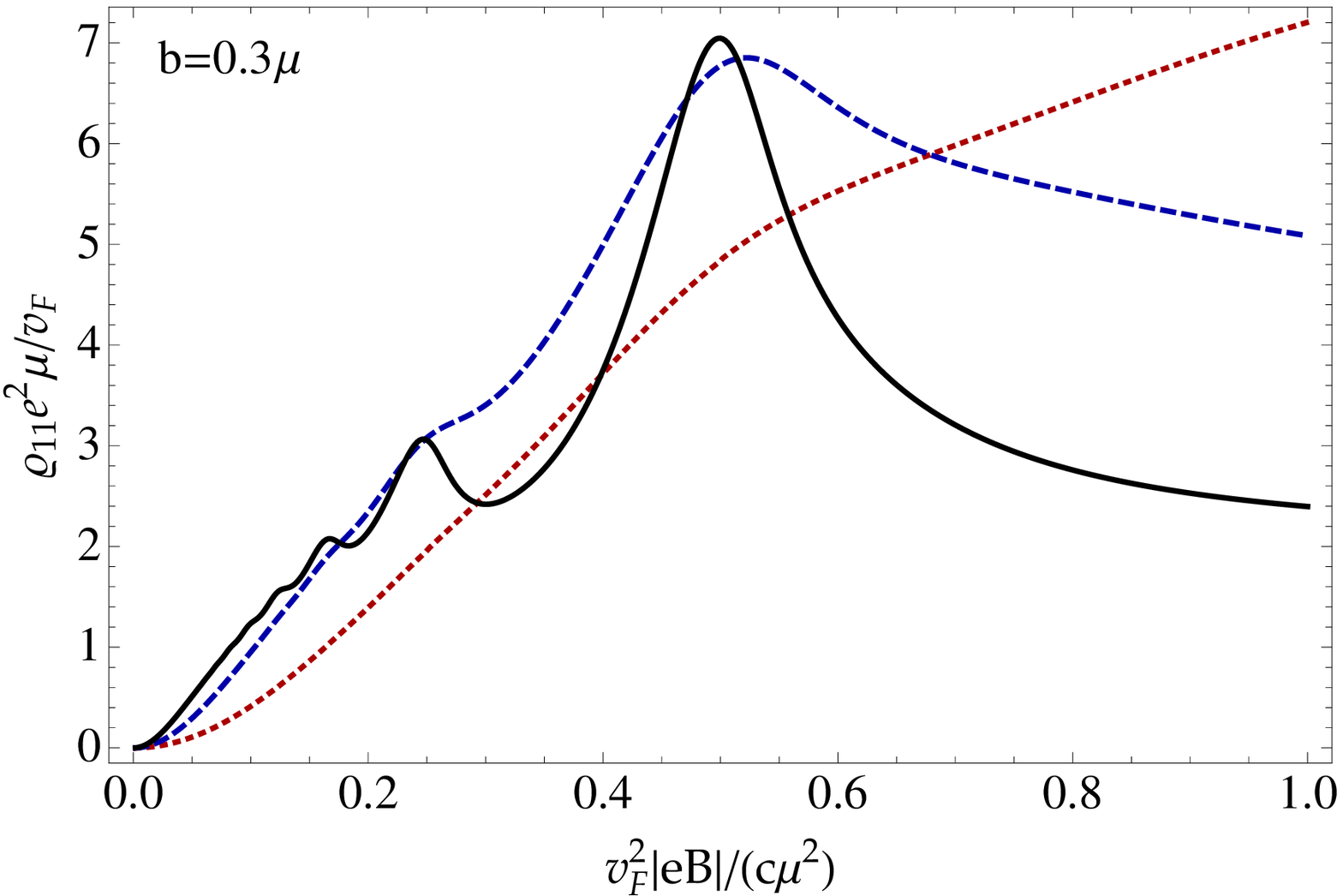}
\hspace{5mm}
\includegraphics[width=.45\textwidth]{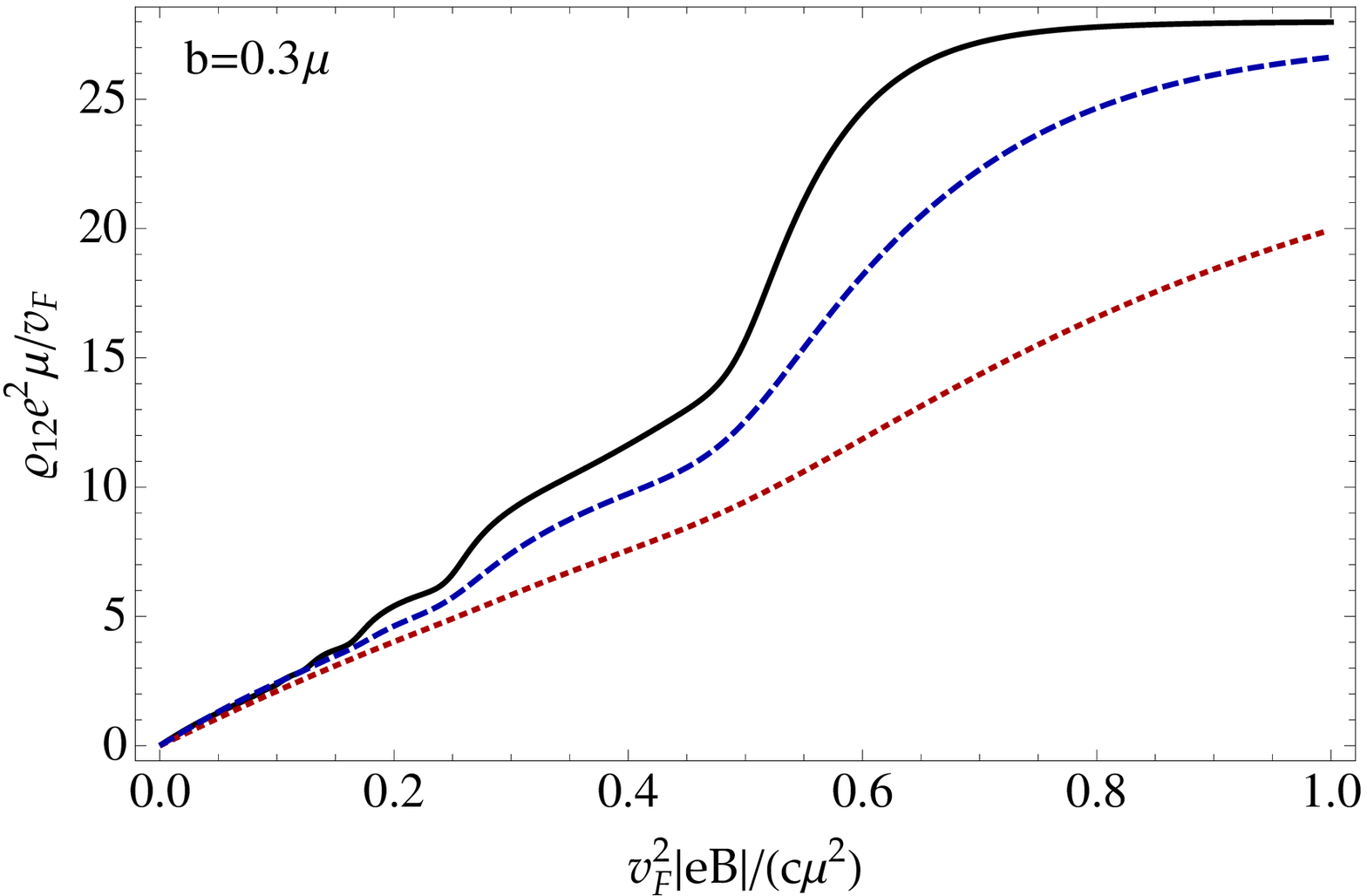}\\
\caption{(Color online) Transverse components of resistivity $\rho_{11}$ and $\rho_{12}$ at zero temperature 
as functions of the magnetic field for $b=0$ (upper panels) and $b=0.3 \mu$ (lower panels).
The quasiparticle width is $\Gamma=0.05\mu$ (black solid line), 
$\Gamma=0.1\mu$ (blue dashed line), and $\Gamma=0.2\mu$ (red dotted line). The sum over 
Landau levels includes $n_{\rm max} =10^{4}$ levels.}
\label{fig:resistivity}
\end{center}
\end{figure}

By making use of the transverse conductivity, we calculate all remaining nonzero 
components of the resistivity tensor, i.e.,
\begin{eqnarray}
\rho_{11} &=& \rho_{22} =\frac{\sigma_{11}}{\sigma_{11}^2+\sigma_{12}^2},\\
\rho_{12} &=&-\rho_{21}=-\frac{\sigma_{12}}{\sigma_{11}^2+\sigma_{12}^2}.
\end{eqnarray}
Using the conductivity results at zero temperature, we calculate $\rho_{11}$ and 
$\rho_{12}$ numerically. The corresponding diagonal and off-diagonal components of 
resistivity are shown as functions of $v^2_F|eB|/(\mu^2c)$ in Fig.~\ref{fig:resistivity} 
for $b=0$ (upper panels) and $b=0.3 \mu$ (lower panels).

\section{Conclusion}
\label{sec:Conclusion}

In this paper, we calculate the longitudinal (along the direction of the magnetic field) and 
transverse (with respect to the direction of the magnetic field) components of the conductivity 
of Dirac and Weyl semimetals in a magnetic field. All calculations are performed in the quantum 
regime by using the Kubo's linear-response theory. We find that all components of conductivity have 
the characteristic Shubnikov-de Haas oscillations as functions of the magnetic field when the 
Landau levels are well resolved (i.e., the quasiparticle widths are sufficiently small). In {\em both} 
Dirac and Weyl semimetals, the magnitudes of the transverse components of conductivity on 
average decrease with increasing the field. We find that {\em both} types of 
semimetals exhibit the regime of negative longitudinal magnetoresistivity at sufficiently large 
magnetic fields and the assumption \cite{Nielsen,Aji,Son} usually made that the decay width 
of quasiparticles in the LLL is much smaller than that in higher Landau levels is not necessary. 
The immediate implication of this fact is that the experimental observation of negative 
longitudinal magnetoresistivity cannot be used alone as an unambiguous signature 
of a Weyl semimetal. 

As our calculations show, the negative magnetoresistivity in the longitudinal 
conductivity occurs solely due to the lowest Landau level.  This contribution has a 
topological origin and is associated with the chiral anomaly. It is also intimately
connected with the dimensional reduction $3 \to 1$ in the dynamics of the LLL
in three-dimensional relativistic-like systems. While the dispersion relation of the 
LLL quasiparticles is independent of the magnetic field, the longitudinal conductivity 
$\sigma_{33}$ grows linearly with the magnetic field because it is proportional to the 
LLL density of states, i.e., $\propto |eB|$. In essence, this growth is the main mechanism 
behind the negative longitudinal magnetoresistivity.

The present results qualitatively agree with the quasiclassical results obtained in 
Refs.~\cite{Nielsen,Aji,Son} using the Boltzmann equation. In general, however,
the quasiclassical results are not sufficient because the quantum corrections due to higher 
Landau levels are quantitatively important in the complete result, especially in the regime 
of moderately strong magnetic fields when a few Landau levels are occupied.

We found that the longitudinal conductivity does not explicitly depend on the value of the shift 
$\mathbf{b}$ between the Weyl nodes. A potential indirect dependence may enter, however,
through the corresponding dependence of the widths of quasiparticles \cite{Nielsen,Aji,Son}. 
This is in contrast to the transverse transport which does reveal an explicit dependence on the 
chiral shift $\mathbf{b}$. Specifically, the off-diagonal transverse component of conductivity 
$\sigma_{12}$ has an anomalous contribution directly proportional to the chiral shift, but 
independent of the temperature, chemical potential and magnetic field. From our analysis,
we see that this anomalous part of conductivity is determined exclusively by the LLL 
quasiparticles. It is also interesting to point out that this contribution has exactly the 
same form as in a Weyl semimetal (with an intrinsic $\mathbf{b}\neq0$) without an external 
magnetic field. It is manifested via the anomalous part of the electric current $\mathbf{j}_{\rm anom}
=e^2/(2\pi^2)\mathbf{b}\times\mathbf{E}$ which is perpendicular to the applied 
electric field \cite{Burkov1,Grushin,ZyuzinBurkov,Franz,Goswami}.

In both Dirac and Weyl semimetals, the chiral shift $\mathbf{b}$ should receive
dynamical corrections proportional to the magnetic field. It would be very interesting to observe 
such corrections experimentally. This is not easy when Landau levels are partially filled and 
an ordinary Hall effect, associated with a nonzero density of charge carriers, is superimposed
over the anomalous Hall conductivity. However, as was demonstrated in Ref.~\cite{Liu} 
in the case of Na$_{3}$Bi, such a problem can be circumvented by tuning the chemical 
potential to the Dirac or Weyl points and, thus, eliminating the contribution due to the ordinary Hall 
effect. This can be done by using surface $K$-doping \cite{Liu}. If this works, it may also allow us to 
observe the dependence of the chiral shift $\mathbf{b}$ on the magnetic field through 
the measurements of the off-diagonal transverse conductivity.

It is also interesting to mention that an experimental observation of a transition 
from a Dirac to Weyl semimetal driven by a magnetic field has been recently reported in 
Ref.~\cite{Kim1307.6990}. By applying moderately strong magnetic fields 
to the Bi$_{1-x}$Sb$_x$ alloy with the antimony concentration of about $x \approx 0.03$ 
(i.e., the regime of a massless Dirac semimetal), the authors observed negative
longitudinal magnetoresistivity and interpreted it as an unambiguous signature of 
the anomaly contribution [see Eq.~(\ref{sigma33LLL})]. As our current study indicates, 
such an observation is indeed the consequence of the anomaly, but not necessarily 
of a Weyl semimetal. In fact, the only direct indication of the Weyl nature of a semimetal
is present in the off-diagonal component of the transverse conductivity $\sigma_{12}$ 
[see Eq.~(\ref{anomaly-contribution})]. Extracting such a contribution from the experimental 
data may be quite challenging, however, because the value of the chiral shift $b$ itself 
is expected to depend on the magnetic field and the density of charge carriers \cite{engineering}.

\acknowledgments

We thank V.~P.~Gusynin for useful remarks. The work of E.V.G. was supported partially by the 
European FP7 program, Grant No.~SIMTECH 246937, SFFR of Ukraine, Grant No.~F53.2/028, 
and the grant STCU\#5716-2 ``Development of Graphene Technologies and Investigation of 
Graphene-based Nanostructures for Nanoelectronics and Optoelectronics". The work of 
V.A.M. was supported by the Natural Sciences and Engineering Research Council of Canada. 
The work of I.A.S. was supported in part by the U.S. National Science Foundation 
under Grant No.~PHY-0969844.
\vspace{3mm}

\appendix

\section{Calculation of traces}
\label{AppA}

In this appendix, we calculate the traces that appear in the definition of the diagonal and off-diagonal 
components of the conductivity [see Eqs.~(\ref{sigma-ii-AA-chi}) and (\ref{sigma-ij-AA-chi}), respectively].
By taking into account that the Dirac structure of the fermion propagator in Eq.~(\ref{full-propagator}) and 
the spectral function (\ref{sp-funct-Gamma}) are the same, we see that all traces have the following 
general structure:
\begin{eqnarray}
 T_{ij} (a,a^\prime) &=& \mbox{tr} \Big\{ 
\gamma^i \left[  (a_0\gamma^0-a_3\gamma^3)({\cal P}_{-}L_{n}-{\cal P}_{+}L_{n-1})
+ c (\mathbf{k}_\perp\cdot\bm{\gamma}_\perp)L_{n-1}^{1}\right] \nonumber\\
&\times&
\gamma^j \left[  (a_0^\prime\gamma^0-a_3^\prime\gamma^3)({\cal P}_{-}L_{n^\prime}-{\cal P}_{+}L_{n^\prime-1})
+ c^\prime(\mathbf{k}_\perp\cdot\bm{\gamma}_\perp)L_{n^\prime-1}^{1}\right] 
 {\cal P}_{5}^{(\chi)}\Big\},
 \label{main-trace}
\end{eqnarray}
where the explicit forms of the coefficients in front of the independent Dirac structures are 
$a_0= E_n^{(\chi)}$, 
$a_0^\prime= E_{n^\prime}^{(\chi)}$, 
$a_3= \lambda v_F (k_3-\chi b)$, 
$a_3^\prime= \lambda^\prime v_F (k_3-\chi b)$, 
$c= 4\lambda v_F$, and 
$c^\prime = 4\lambda^\prime v_F$.

The traces are straightforward to calculate. The results read as
\begin{eqnarray}
T_{11}  &=& -\left(a_0 a_0^\prime-a_3 a_3^\prime\right)\left(L_{n-1} L_{n^\prime}+L_{n} L_{n^\prime-1}\right)
-s_\perp\chi\left(a_0a_3^\prime-a_0^\prime a_3\right)\left(L_{n-1} L_{n^\prime}-L_{n} L_{n^\prime-1}\right)  
\nonumber\\
&&
+2c c^\prime\left(k_1^2-k_2^2\right) L_{n-1}^{1} L_{n^\prime-1}^{1}, \\
T_{22}  &=&-\left(a_0 a_0^\prime-a_3 a_3^\prime\right)\left(L_{n-1} L_{n^\prime}+L_{n} L_{n^\prime-1}\right)
-s_\perp\chi\left(a_0a_3^\prime-a_0^\prime a_3\right)\left(L_{n-1} L_{n^\prime}-L_{n} L_{n^\prime-1}\right) 
\nonumber\\
&&
-2c c^\prime\left(k_1^2-k_2^2\right) L_{n-1}^{1} L_{n^\prime-1}^{1},\\
T_{12}  &=& i s_\perp\left(a_0 a_0^\prime-a_3 a_3^\prime \right)\left(L_{n-1} L_{n^\prime}-L_{n} L_{n^\prime-1} \right)
+i \chi\left(a_0 a_3^\prime - a_0^\prime a_3\right)\left(L_{n-1} L_{n^\prime}+L_{n} L_{n^\prime-1} \right)
\nonumber\\
&&
+4 c c^\prime k_1k_2L_{n-1}^{1} L_{n^\prime-1}^{1},\\
T_{21}  &=& - i s_\perp\left(a_0 a_0^\prime-a_3 a_3^\prime \right)\left(L_{n-1} L_{n^\prime}-L_{n} L_{n^\prime-1} \right)
-i \chi\left(a_0 a_3^\prime - a_0^\prime a_3\right)\left(L_{n-1} L_{n^\prime}+L_{n} L_{n^\prime-1} \right)
\nonumber\\
&&
+4 c c^\prime k_1k_2 L_{n-1}^{1} L_{n^\prime-1}^{1},\\
T_{33}  &=& \left(a_0 a_0^\prime+a_3 a_3^\prime\right)\left(L_{n} L_{n^\prime}+L_{n-1} L_{n^\prime-1}\right)
+s_\perp\chi\left(a_0^\prime a_3+a_0 a_3^\prime\right)\left(L_{n} L_{n^\prime}-L_{n-1} L_{n^\prime-1}\right)
\nonumber\\
&&
-2c c^\prime k_\perp^2 L_{n-1}^{1} L_{n^\prime-1}^{1},
\end{eqnarray}
\begin{eqnarray}
T_{13}  &=& -k_1\Big[a_3 c^\prime \left( L_{n} - L_{n-1}  \right) L_{n^\prime-1}^{1} 
                               + a_3^\prime c \left( L_{n^\prime} - L_{n^\prime-1}  \right) L_{n-1}^{1} 
                               +s_\perp\chi a_0 c^\prime \left( L_{n} + L_{n-1}  \right) L_{n^\prime-1}^{1} 
\nonumber\\
&&
                               + s_\perp\chi a_0^\prime c \left( L_{n^\prime} + L_{n^\prime-1}  \right) L_{n-1}^{1} 
                               \Big] 
                                - i k_2\Big[s_\perp a_3 c^\prime \left( L_{n} + L_{n-1}  \right) L_{n^\prime-1}^{1} 
                               - s_\perp a_3^\prime c \left( L_{n^\prime} + L_{n^\prime-1}  \right) L_{n-1}^{1} 
\nonumber\\
&&
                               +\chi a_0 c^\prime \left( L_{n} - L_{n-1}  \right) L_{n^\prime-1}^{1} 
                               - \chi a_0^\prime c \left( L_{n^\prime} - L_{n^\prime-1}  \right) L_{n-1}^{1} 
                               \Big] ,\\
T_{31}  &=& -k_1\Big[a_3 c^\prime \left( L_{n} - L_{n-1}  \right) L_{n^\prime-1}^{1} 
                               + a_3^\prime c \left( L_{n^\prime} - L_{n^\prime-1}  \right) L_{n-1}^{1} 
                               +s_\perp\chi a_0 c^\prime \left( L_{n} + L_{n-1}  \right) L_{n^\prime-1}^{1} 
\nonumber\\
&&
                               + s_\perp\chi a_0^\prime c \left( L_{n^\prime} + L_{n^\prime-1}  \right) L_{n-1}^{1} 
                               \Big]
                               + i k_2\Big[s_\perp a_3 c^\prime \left( L_{n} + L_{n-1}  \right) L_{n^\prime-1}^{1} 
                               - s_\perp a_3^\prime c \left( L_{n^\prime} + L_{n^\prime-1}  \right) L_{n-1}^{1} 
\nonumber\\
&&
                               +\chi a_0 c^\prime \left( L_{n} - L_{n-1}  \right) L_{n^\prime-1}^{1} 
                               - \chi a_0^\prime c \left( L_{n^\prime} - L_{n^\prime-1}  \right) L_{n-1}^{1} 
                               \Big] ,\\
T_{23}  &=&   i k_1\Big[s_\perp a_3 c^\prime \left( L_{n} + L_{n-1}  \right) L_{n^\prime-1}^{1} 
                               - s_\perp a_3^\prime c \left( L_{n^\prime} + L_{n^\prime-1}  \right) L_{n-1}^{1} 
                               +\chi a_0 c^\prime \left( L_{n} - L_{n-1}  \right) L_{n^\prime-1}^{1} 
\nonumber\\
&&
                               - \chi a_0^\prime c \left( L_{n^\prime} - L_{n^\prime-1}  \right) L_{n-1}^{1} 
                               \Big]\
                               - k_2\Big[a_3 c^\prime \left( L_{n} - L_{n-1}  \right) L_{n^\prime-1}^{1} 
                               + a_3^\prime c \left( L_{n^\prime} - L_{n^\prime-1}  \right) L_{n-1}^{1} 
\nonumber\\
&&
                               +s_\perp \chi a_0 c^\prime \left( L_{n} + L_{n-1}  \right) L_{n^\prime-1}^{1} 
                               + s_\perp \chi a_0^\prime c \left( L_{n^\prime} + L_{n^\prime-1}  \right) L_{n-1}^{1} 
                               \Big] , \\
T_{32}  &=&  - i k_1\Big[s_\perp a_3 c^\prime \left( L_{n} + L_{n-1}  \right) L_{n^\prime-1}^{1} 
                               - s_\perp a_3^\prime c \left( L_{n^\prime} + L_{n^\prime-1}  \right) L_{n-1}^{1} 
                               +\chi a_0 c^\prime \left( L_{n} - L_{n-1}  \right) L_{n^\prime-1}^{1} 
\nonumber\\
&&
                               - \chi a_0^\prime c \left( L_{n^\prime} - L_{n^\prime-1}  \right) L_{n-1}^{1} 
                               \Big]
                                - k_2\Big[a_3 c^\prime \left( L_{n} - L_{n-1}  \right) L_{n^\prime-1}^{1} 
                               + a_3^\prime c \left( L_{n^\prime} - L_{n^\prime-1}  \right) L_{n-1}^{1} 
\nonumber\\
&&
                               +s_\perp \chi a_0 c^\prime \left( L_{n} + L_{n-1}  \right) L_{n^\prime-1}^{1} 
                               + s_\perp \chi a_0^\prime c \left( L_{n^\prime} + L_{n^\prime-1}  \right) L_{n-1}^{1} 
                               \Big].
\end{eqnarray}
After the integration over the transverse momenta in the expression for the conductivity, the terms 
linear in $k_1$ and $k_2$, as well as the terms proportional to $k_1^2-k_2^2$ will vanish. This is 
equivalent to replacing the traces with expressions averaged over the transverse directions, i.e., 
$T_{ij} \to \tilde{T}_{ij}$, where 
\begin{eqnarray}
\tilde{T}_{11}  &=&  -\left(a_0 a_0^\prime-a_3 a_3^\prime\right)\left(L_{n-1} L_{n^\prime}+L_{n} L_{n^\prime-1}\right)
-s_\perp\chi\left(a_0a_3^\prime-a_0^\prime a_3\right)\left(L_{n-1} L_{n^\prime}-L_{n} L_{n^\prime-1}\right), \\
\tilde{T}_{12}  &=&   i s_\perp\left(a_0 a_0^\prime-a_3 a_3^\prime \right)\left(L_{n-1} L_{n^\prime}-L_{n} L_{n^\prime-1} \right)
+i \chi\left(a_0 a_3^\prime - a_0^\prime a_3\right)\left(L_{n-1} L_{n^\prime}+L_{n} L_{n^\prime-1} \right), \\
\tilde{T}_{33} &=& \left(a_0 a_0^\prime+a_3 a_3^\prime\right)\left(L_{n} L_{n^\prime}+L_{n-1} L_{n^\prime-1}\right)
+s_\perp\chi\left(a_0^\prime a_3+a_0 a_3^\prime\right)\left(L_{n} L_{n^\prime}-L_{n-1} L_{n^\prime-1}\right)
-2c c^\prime k_\perp^2 L_{n-1}^{1} L_{n^\prime-1}^{1},
\end{eqnarray}
as well as $\tilde{T}_{22} \equiv \tilde{T}_{11} $ and $\tilde{T}_{21} \equiv - \tilde{T}_{12}$. The other 
off-diagonal components vanish, i.e., $\tilde{T}_{13}=\tilde{T}_{31}=\tilde{T}_{23}=\tilde{T}_{32}=0$.

The dependence on the transverse momenta in the resulting traces $\tilde{T}_{ij}$ enters 
only via the Laguerre polynomials. Therefore, after these results are substituted into the expressions 
for the conductivity in Eqs.~(\ref{sigma-ii-AA-chi}) and (\ref{sigma-ij-AA-chi}), the subsequent 
integration over $\mathbf{k}_\perp$ can be easily performed. Indeed, by making use of the 
orthogonality of the Laguerre polynomials, 
\begin{equation}
\int_{0}^{\infty} x^\alpha e^{-x} L_{n}^{(\alpha)}(x)L_{m}^{(\alpha)}(x) dx = 
\frac{\Gamma(m+1+\alpha)}{n!}\delta_{n,m},
\end{equation}
we derive the following integration results: 
\begin{eqnarray}
X_{11} = \int \frac{d^2 \mathbf{k}_\perp}{(2\pi)^2} e^{-2k_\perp^2 l^{2}} \tilde{T}_{11} &=& 
-\frac{a_0 a_0^\prime-a_3 a_3^\prime}{8\pi l^2}\left(\delta_{n-1,n^\prime}+\delta_{n,n^\prime-1}\right)
-s_\perp\chi\frac{a_0a_3^\prime-a_0^\prime a_3}{8\pi l^2 }\left(\delta_{n-1,n^\prime}-\delta_{n,n^\prime-1}\right),
\label{X11} \\
X_{12} = \int \frac{d^2 \mathbf{k}_\perp}{(2\pi)^2}  e^{-2k_\perp^2 l^{2}} \tilde{T}_{12} &=& 
 i s_\perp\frac{a_0 a_0^\prime-a_3 a_3^\prime }{8\pi l^2 }\left(\delta_{n-1,n^\prime}-\delta_{n,n^\prime-1} \right)
+i \chi\frac{a_0 a_3^\prime - a_0^\prime a_3}{8\pi l^2 }\left(\delta_{n-1,n^\prime}+\delta_{n,n^\prime-1} \right),
\label{X12} \\
X_{33} = \int \frac{d^2 \mathbf{k}_\perp}{(2\pi)^2} e^{-2k_\perp^2 l^{2}}  \tilde{T}_{33} &=& 
\frac{a_0 a_0^\prime+a_3 a_3^\prime}{8\pi l^2 }\left(\delta_{n,n^\prime}+\delta_{n-1,n^\prime-1}\right)
+s_\perp\chi\frac{a_0^\prime a_3+a_0 a_3^\prime}{8\pi l^2 }\left(\delta_{n,n^\prime}-\delta_{n-1,n^\prime-1}\right)
\nonumber\\
&&
-\frac{n c c^\prime }{8\pi l^4}\delta_{n-1,n^\prime-1}.
\label{X33} 
\end{eqnarray}
We use these results in the main text when calculating the transverse and longitudinal components 
of the conductivity tensor, i.e.,
$\sigma_{11}=\sigma_{22}$,
$\sigma_{12}=-\sigma_{12}$, and 
$\sigma_{33}$. 

\section{Key details in derivation of conductivity}
\label{AppB}

\subsection{Longitudinal conductivity}

By making use of the definition in Eq.~(\ref{sigma-ii-AA-chi}) as well as the result for the trace in Eq.~(\ref{X33}), 
we derive the following expression for the longitudinal component of the conductivity:
\begin{eqnarray}
\sigma_{33} &=& \frac{e^2 v_F^2 }{2^6\pi^3 l^2 T}
\sum_{\chi} \sum_{\lambda,\lambda^\prime}\sum_{n=0}^{\infty}
\int \frac{d \omega d k^3 }{\cosh^2\frac{\omega-\mu}{2T}}
 \frac{\Gamma_n^2}{\left[\left(\omega -\lambda E_n^{(\chi)} \right)^2+\Gamma_n^2\right]
\left[ \left(\omega -\lambda^\prime E_{n}^{(\chi)} \right)^2+\Gamma_{n}^2\right]} \nonumber \\
&\times&  
\left(1+\lambda s_\perp\chi v_F\frac{k^3-\chi b}{E_n^{(\chi)} }\right)
\left(1+\lambda^\prime s_\perp \chi  v_F\frac{k^3-\chi b}{E_{n}^{(\chi)}}\right)\nonumber \\
&+&  \frac{e^2 v_F^2 }{2^6\pi^3 l^2 T}
\sum_{\chi} \sum_{\lambda,\lambda^\prime}\sum_{n=1}^{\infty}
\int \frac{d \omega d k^3 }{\cosh^2\frac{\omega-\mu}{2T}}
 \frac{\Gamma_n^2}{\left[\left(\omega -\lambda E_n^{(\chi)} \right)^2+\Gamma_n^2\right]
\left[ \left(\omega -\lambda^\prime E_{n}^{(\chi)} \right)^2+\Gamma_{n}^2\right]} \nonumber \\
&\times&  \left[
\left(1-\lambda s_\perp\chi v_F\frac{k^3-\chi b}{E_n^{(\chi)} }\right)
\left(1-\lambda^\prime s_\perp \chi  v_F\frac{k^3-\chi b}{E_{n}^{(\chi)}}\right)
-\frac{4 v_F^2\lambda \lambda^\prime n}{E_{n}^{(\chi)} E_{n^\prime}^{(\chi)} l^2}\right].
\end{eqnarray}
After performing the sum over $\lambda$ and $\lambda^\prime$, we can rewrite this result in
a more convenient form given by Eq.~(\ref{conductivity-longitudinal}) in the main text. 

Because of a qualitatively different role that the lowest and higher Landau levels play in the 
magnetoresistance, we find it convenient to separate the two contributions. The corresponding 
expressions for $\sigma_{33}^{\rm (LLL)}$ and $\sigma_{33}^{\rm (HLL)} $ are given in 
Eqs.~(\ref{sigma33LLL}) and (\ref{sigma33HLL}) in the main text. While the former takes 
a very simple analytical form, the latter is much more complicated. Some details of its 
analysis are presented here. As stated in the main text, the integration over $k_3$ in the 
expression for $\sigma_{33}^{\rm (HLL)}$ can be performed analytically. The corresponding 
result reads
\begin{eqnarray}
\sigma_{33}^{\rm (HLL)} &=&\frac{e^2 v_F}{4\sqrt{2}\pi^2 l^2 T}
\sum_{n=1}^{\infty} 
\int \frac{d \omega}{\cosh^2\frac{\omega-\mu}{2T}}
\frac{\Gamma_n^2}{
\sqrt{\sqrt{\left(2n \epsilon_{L}^2+\Gamma_n^2-\omega^2\right)^2 + 4 \omega^2 \Gamma_n^2}+2n \epsilon_{L}^2+\Gamma_n^2-\omega^2}}
\nonumber\\
&&\frac{1}{\sqrt{\left(2n \epsilon_{L}^2+\Gamma_n^2-\omega^2\right)^2 + 4 \omega^2 \Gamma_n^2}}
\Bigg[1+\frac{\omega^2}{\sqrt{\left(2n \epsilon_{L}^2+\Gamma_n^2-\omega^2\right)^2 + 4 \omega^2 \Gamma_n^2}+2n \epsilon_{L}^2+\Gamma_n^2-\omega^2} \nonumber\\
&&\times \left(1+\frac{n \epsilon_{L}^2\left[3\sqrt{\left(2n \epsilon_{L}^2+\Gamma_n^2-\omega^2\right)^2 + 4 \omega^2 \Gamma_n^2}+2\left(2n \epsilon_{L}^2+\Gamma_n^2-\omega^2\right)\right]}{\left(2n \epsilon_{L}^2+\Gamma_n^2-\omega^2\right)^2 + 4 \omega^2 \Gamma_n^2}\right)
\Bigg].
\label{sigma33HLL_K3}
\end{eqnarray}
In the zero temperature limit, additionally the remaining integration over $\omega$ 
can be performed as well. The corresponding result is given by
\begin{eqnarray}
\sigma_{33,T\to 0}^{\rm (HLL)} &=&\frac{e^2 v_F}{\sqrt{2} \pi^2 l^2}
\sum_{n=1}^{\infty} \frac{\Gamma^2}{\sqrt{\sqrt{\left(2n \epsilon_{L}^2+\Gamma^2-\mu^2\right)^2 + 4 \mu^2 \Gamma^2}+2n \epsilon_{L}^2+\Gamma^2-\mu^2}}
\nonumber\\
&&\frac{1}{\sqrt{\left(2n \epsilon_{L}^2+\Gamma^2-\mu^2\right)^2 + 4 \mu^2 \Gamma^2}}
\Bigg[1+\frac{\mu^2}{\sqrt{\left(2n \epsilon_{L}^2+\Gamma^2-\mu^2\right)^2 + 4 \mu^2 \Gamma^2}+2n \epsilon_{L}^2+\Gamma^2-\mu^2} 
\nonumber\\
&&\times \left(1+\frac{n \epsilon_{L}^2\left[3\sqrt{\left(2n \epsilon_{L}^2+\Gamma^2-\mu^2\right)^2 + 4 \mu^2 \Gamma^2}+2\left(2n 
\epsilon_{L}^2+\Gamma^2-\mu^2\right)\right]}{\left(2n \epsilon_{L}^2+\Gamma^2-\mu^2\right)^2 + 4 \mu^2 \Gamma^2}\right)
\Bigg],
\label{sigma33HLL_omega}
\end{eqnarray}
where, for simplicity, we took $\Gamma_n\equiv \Gamma$  in all higher Landau levels.

\subsection{Transverse conductivity}

By making use of the definition in Eq.~(\ref{sigma-ii-AA-chi}) as well as the result for the trace in Eq.~(\ref{X11}), 
we derive the following expression for the diagonal component of the transverse conductivity:
\begin{eqnarray}
\sigma_{11} &=& \frac{e^2 v_F^2 }{2^5\pi^3 l^2 T}
\sum_{\chi} \sum_{\lambda,\lambda^\prime}\sum_{n=0}^{\infty}
\int \frac{d \omega d k_3 }{\cosh^2\frac{\omega-\mu}{2T}}
 \frac{\Gamma_{n+1} \Gamma_{n}}{\left[\left(\omega -\lambda E_{n+1}^{(\chi)} \right)^2+\Gamma_{n+1}^2\right]
\left[ \left(\omega -\lambda^\prime E_{n}^{(\chi)} \right)^2+\Gamma_{n}^2\right]} \nonumber \\
&\times&
\left(1-\lambda s_\perp\chi v_F\frac{k_3-\chi b}{E_{n+1}^{(\chi)} }\right)
\left(1+\lambda^\prime s_\perp \chi  v_F\frac{k_3-\chi b}{E_{n}^{(\chi)}}\right).
\end{eqnarray}
By taking into account that the momentum integral is convergent, we can make the shift of 
the integration variable $k_3  \to  k_{\rm new}^3\equiv k_3-\chi b$. 
Then the integrand does not depend on $b$ and we find
\begin{eqnarray}
\sigma_{11} &=& \frac{e^2 v_F^2 }{2^4\pi^3 l^2 T}
\sum_{\lambda,\lambda^\prime}\sum_{n=0}^{\infty}
\int \frac{d \omega d k_3 }{\cosh^2\frac{\omega-\mu}{2T}}
 \frac{\Gamma_{n+1} \Gamma_{n}}{\left[\left(\omega -\lambda E_{n+1} \right)^2+\Gamma_{n+1}^2\right]
\left[ \left(\omega -\lambda^\prime E_{n} \right)^2+\Gamma_{n}^2\right]}  
\left(1-\lambda \lambda^\prime \frac{(v_F k_3)^2}{E_{n+1} E_{n}}\right).
\end{eqnarray}
After calculating the sum over $\lambda$ and $\lambda^\prime$, we will obtain the result presented 
in Eq.~(\ref{sigma11_text}) in the main text. 

In the limit of zero temperature, both integrations over $\omega$ and $k_3$ in the expression for the
diagonal component of the transverse conductivity can be performed analytically. The corresponding 
result reads
\begin{eqnarray}
\sigma_{11} &=& \frac{e^2 \epsilon_L^2 \Gamma^2}{ 2 \sqrt{2} \pi^2 v_F }
\sum_{n=0}^{\infty}\Bigg\{
\frac{1}{ \sqrt{\left(2n\epsilon_{L}^2+\Gamma^2 -\mu^2\right)^2+4\mu^2 \Gamma^2}\sqrt{2n\epsilon_{L}^2+\Gamma^2 
-\mu^2+\sqrt{\left(2n\epsilon_{L}^2+\Gamma^2 -\mu^2\right)^2+4\mu^2 \Gamma^2}}}\nonumber\\
&&
+\frac{1}{ \sqrt{\left(2(n+1)\epsilon_{L}^2+\Gamma^2 -\mu^2\right)^2+4\mu^2 \Gamma^2}\sqrt{2(n+1)\epsilon_{L}^2+\Gamma^2 
-\mu^2+\sqrt{\left(2(n+1)\epsilon_{L}^2+\Gamma^2 -\mu^2\right)^2+4\mu^2 \Gamma^2}}}
\nonumber\\
&& 
+\frac{ 2(2n+1)\mu^2 -\epsilon_{L}^2   }
{ \epsilon_{L}^4+4 \mu^2 \Gamma^2 }
\left[-\frac{1}{\sqrt{2n\epsilon_{L}^2+\Gamma^2 -\mu^2+\sqrt{\left(2n\epsilon_{L}^2+\Gamma^2 -\mu^2\right)^2+4\mu^2 \Gamma^2}}}\right.
\nonumber\\
&& \left.
-\frac{(2n-1)\epsilon_{L}^2+\Gamma^2 -\mu^2 }
{\sqrt{\left(2n\epsilon_{L}^2+\Gamma^2 -\mu^2\right)^2+4\mu^2 \Gamma^2}\sqrt{2n\epsilon_{L}^2+\Gamma^2 
-\mu^2+\sqrt{\left(2n\epsilon_{L}^2+\Gamma^2 -\mu^2\right)^2+4\mu^2 \Gamma^2}}} \right.\nonumber\\
&& \left.
+\frac{1}{\sqrt{2(n+1)\epsilon_{L}^2+\Gamma^2 -\mu^2+\sqrt{\left(2(n+1)\epsilon_{L}^2+\Gamma^2 -\mu^2\right)^2+4\mu^2 \Gamma^2}}}\right.
\nonumber\\
&& \left.
+\frac{(2n+3)\epsilon_{L}^2+\Gamma^2 -\mu^2 }{\sqrt{\left(2(n+1)\epsilon_{L}^2+\Gamma^2 -\mu^2\right)^2+4\mu^2 
\Gamma^2}\sqrt{2(n+1)\epsilon_{L}^2+\Gamma^2 -\mu^2+\sqrt{\left(2(n+1)\epsilon_{L}^2+\Gamma^2 -\mu^2\right)^2+4\mu^2 \Gamma^2}}}
\right]
\Bigg\},
\label{sigma11}
\end{eqnarray}
where, for simplicity, we took $\Gamma_n\equiv \Gamma$  in all Landau levels.
Note that the function in the sum over Landau levels is $\propto n^{-3/2}$ when $n\to \infty$ and, therefore, 
the sum converges quite fast and is easily calculated by numerical methods.


\begin{thebibliography}{99}
\bibitem{footnote1}
In this context, negative magnetoresistivity is realized when the resistivity (conductivity)
decreases (increases) with the magnetic field.
\end{thebibliography}
\end{document}